\documentstyle[floats,prd,aps,preprint,epsf]{revtex}

\draft
\begin{document}
%\twocolumn[\hsize\textwidth\columnwidth\hsize\csname
%@twocolumnfalse\endcsname

%
\makeatletter
\@ifundefined{lesssim}{\def\lesssim{\mathrel{\mathpalette\vereq<}}}{}
\@ifundefined{gtrsim}{\def\gtrsim{\mathrel{\mathpalette\vereq>}}}{}
\def\vereq#1#2{\lower3pt\vbox{\baselineskip1.5pt \lineskip1.5pt
\ialign{$\m@th#1\hfill##\hfil$\crcr#2\crcr\sim\crcr}}}
\makeatother
\newcommand{\beq}{\begin{equation}}
\newcommand{\eeq}{\end{equation}}
\newcommand{\beqn}{\begin{eqnarray}}
\newcommand{\eeqn}{\end{eqnarray}}
\newcommand{\pa}{\partial}
\newcommand{\vp}{\varphi}
\newcommand{\varep}{\varepsilon}
\def\zero{\hbox{$_{(0)}$}}
\def\bL{\hbox{$\,{\cal L}\!\!\!$--}}
\def\bI{\hbox{$\,I\!\!\!$--}}

\begin{center}
{\large\bf{Gravitational waves from axisymmetrically 
oscillating neutron stars in general relativistic simulations
}}
~\\
~\\
Masaru Shibata and Yu-ichirou Sekiguchi\\
{\em Graduate School of Arts and 
Sciences, University of Tokyo, Tokyo, 153-8902, Japan}\\
\end{center}
\begin{abstract}
Gravitational waves from oscillating neutron stars 
in axial symmetry are studied performing numerical 
simulations in full general relativity. Neutron stars are
modeled by a polytropic equation of state for simplicity. 
A gauge-invariant wave extraction method as well as a 
quadrupole formula are adopted for computation of gravitational waves. 
It is found that the gauge-invariant variables systematically contain 
numerical errors generated near the outer boundaries in the present 
axisymmetric computation. We clarify their origin, and illustrate it 
possible to eliminate the dominant part of the systematic errors. 
The best corrected waveforms for oscillating and rotating stars currently 
contain errors of magnitude $\sim 10^{-3}$ in the local wave zone. 
Comparing the waveforms obtained by the gauge-invariant technique 
with those by the quadrupole formula, it is shown 
that the quadrupole formula yields approximate gravitational waveforms
besides a systematic underestimation of the amplitude of $O(M/R)$ 
where $M$ and $R$ denote the mass and the radius of neutron stars.
However, the wave phase and modulation of the amplitude can be
computed accurately. This indicates that the quadrupole formula is a
useful tool for studying gravitational waves from rotating stellar
core collapse to a neutron star in fully general 
relativistic simulations. Properties of the gravitational waveforms from 
the oscillating and rigidly rotating neutron stars are also addressed
paying attention to the oscillation associated with fundamental modes. 
\end{abstract}
%\vskip2pc]

\section{Introduction}

One of the most important roles of numerical simulations in general
relativity is to predict gravitational waveforms emitted by 
general relativistic and dynamical astrophysical phenomena. 
Rotating stellar core collapse and nonspherical oscillation of neutron stars 
are among the possible sources of gravitational waves. Therefore, 
fully general relativistic numerical simulation for them 
is an important subject in this field \cite{HD}. 

To date, there has been no systematic work for computation of 
gravitational waves from rotating stellar core collapse to a neutron star 
in fully general relativistic simulation (but see \cite{Siebel}). 
The gravitational waveforms have been computed only in the Newtonian gravity 
\cite{Newton,Newton1,Newton2,Newton3,Newton4,Newton5,Newton6} 
or in an approximate general relativistic 
gravity \cite{HD} using the so-called conformal flatness 
approximation (or Isenberg-Wilson-Mathews approximation). 
As demonstrated in \cite{HD}, general relativistic effects modify 
the evolution of the collapse and emitted gravitational waveforms
significantly. 
Thus, the simulation in full general relativity appears to be the best 
approach for accurate computation of gravitational waves. 

In the case that
the progenitor of the core collapse is not very rapidly rotating, 
nonaxisymmetric instabilities do not set in and, hence, 
the collapse will proceed in an axisymmetric manner. 
In such a collapse, the amplitude of gravitational waves
measured in a local wave zone at $r \approx \lambda$ where
$\lambda$ denotes the gravitational wave length will be 
smaller, by two or three orders of magnitude, 
than that in highly nonaxisymmetric phenomena such as 
mergers of binary neutron stars and black holes. 
The amplitude of gravitational waves from an oscillating neutron star is
also likely to be small due to its small nonspherical deformation. 
Technically, it is not easy to extract gravitational waves of small
amplitude from metric computed in numerical simulations, in which 
a numerical noise is in general contained. 
The numerical noise is generated due to the following reasons: 

Gravitational waves are usually extracted from the metric 
in the wave zone in general relativistic simulations. 
Although they should be extracted at the 
null infinity, the outer boundaries of computational domain are 
located at a finite radius whenever the 3+1 formalisms are adopted. 
Thus, the outer boundary conditions are imposed at finite radii 
and in general they are approximate conditions. 
As a result, a small numerical error may be excited around the outer 
boundaries. Here, the possible candidates of the numerical error are  
unphysical nonwave modes, spurious gauge modes, 
back reflections at the outer boundaries, and roundoff errors. 

%%These can contaminate the gravitational waveforms numerically computed. 
%%To remove such errors, as a first step, it is necessary to clarify 
%%what may prevent accurate computation of gravitational waves. 

In this paper, we study gravitational waves from oscillating 
neutron stars in axial symmetry. 
Neutron stars in equilibrium are simply modeled by $n=1$ polytropes.
Oscillations of neutron stars are followed 
by axisymmetric numerical simulations in full general relativity. 
Gravitational waves are extracted using a gauge-invariant wave 
extraction technique. The gauge-invariant variables are not 
contaminated by gauge modes and, hence, we can focus on other error 
sources using this variables. We also adopt a quadrupole formula 
for approximately computing gravitational waveforms 
to clarify its validity. 

This work was planned from the following four motivations. 
The first one is to specify the error sources contained in 
the gauge-invariant variables extracted in the local wave zone. 
As mentioned above, they could be contaminated by nonwave
components and numerical errors. In particular, it is important 
to specify systematic error components 
contained in the gauge-invariant variables since 
as indicated in Sec. IV, the systematic errors may be eliminated
at least partly if their origin is clarified.  

The second motivation is to understand
how large computational domains are needed to extract 
gravitational waveforms within $\sim 10$\% error. 
Since the gauge-invariant variables are extracted at finite 
radii, gravitational waveforms (in particular the amplitude) are
slightly different from the asymptotic ones. It is important to 
clarify how magnitude of the error depends on the 
radius at which we impose the outer boundary conditions and on 
the radius at which we extract gravitational waves. 
A similar study was carried out about 15 years ago 
by Abrahams and Evans \cite{AE}. However, they 
were interested only in specific gauge conditions 
which were often used in axisymmetric numerical simulations in
general relativity at that time.
Moreover, the simulations were carried out only for non-rotating stars. 
In this paper, we adopt a different gauge condition often used nowadays
in three-dimensional simulations, and report numerical results 
both for nonrotating and rotating stars. 

The third motivation, in which we are most interested in the present
study, is to investigate validity of a quadrupole formula 
in fully general relativistic simulations. 
For computation of gravitational waves generated by 
oscillations of gravitational field such as 
quasinormal mode ringings of black holes, 
quadrupole formulas cannot work. However, 
in rotating stellar core collapse to a neutron star 
and in oscillating neutron stars in which 
gravitational waves are generated mainly by matter motions,
quadrupole formulas may be able to yield an accurate waveform. 
This method can be applied much more easily 
than geometrical methods in which gravitational
waves are extracted from metric in the wave zone.
Thus, a quadrupole formula which can yield high-quality 
approximate waveforms will be a robust method for computing
gravitational waves of small amplitude from a noisy numerical data set. 
Note that a similar work has been already done by Siebel et 
al. \cite{Siebel0,Siebel} in a null-cone formulation.
We here carry out the similar study for a 3+1 approach. 

The last motivation is to understand the properties of oscillations
of rotating neutron stars. During 
rotating stellar core collapse, gravitational waves associated with
oscillations of a formed protoneutron star are
likely to be emitted (see, e.g., \cite{HD}). 
From the study for oscillating and rotating neutron stars,
we will be able to understand what oscillation modes are relevant for
the emission of gravitational waves during core collapses. 
We here pay attention to two fundamental oscillation modes
(quasiradial and quadrupole $p$ modes of no node for the
density perturbation) which are candidates for the dominant modes
in the oscillating and rotating stars formed 
after the collapse. 

This paper is organized as follows. 
In Sec. II, our numerical implementations for 
axisymmetric general relativistic simulation are briefly reviewed. 
In Sec. III, the gauge-invariant wave extraction technique 
and the quadrupole formula adopted in the present work are described. 
Sec. IV presents numerical results of gravitational waveforms 
emitted from oscillating neutron stars. 
The simulations were performed both for nonrotating and 
rotating neutron stars using an 
axisymmetric code recently developed \cite{S2002}. 
Sec. IV is devoted to a summary. 
Throughout this paper, we adopt the geometrical units 
in which $G=c=1$ where $G$ and $c$ are the 
gravitational constant and the speed of light, respectively.

\section{Numerical implementation}

\subsection{Summary of formulation}

We performed fully general relativistic simulations 
in axial symmetry using the same formulation as that 
in~\cite{S2002}, to which the readers may refer for details and basic 
equations. The fundamental variables for hydrodynamics are: 
\beqn 
\rho &&:{\rm rest~ mass~ density},\nonumber \\
\varep &&: {\rm specific~ internal~ energy}, \nonumber \\
P &&:{\rm pressure}, \nonumber \\
u^{\mu} &&: {\rm four~ velocity}, \nonumber \\
v^i &&={dx^i \over dt}={u^i \over u^t},
\eeqn
where subscripts $i, j, k, \cdots$ denote $x, y$, and $z$, and
$\mu$ the spacetime components. In addition, 
we define a weighted density $\rho_*(=\rho \alpha u^t e^{6\phi})$
and a weighted four-velocity $\hat u_i (= (1+\varepsilon+P/\rho)u_i)$.
From these variables, the total baryon rest-mass and angular momentum 
of system, which are conserved quantities in axisymmetric
spacetimes, can be defined as 
\beqn
M_*&=&\int d^3 x \rho_*, \\
J  &=&\int d^3 x \rho_*\hat u_{\varphi}. 
\eeqn
General relativistic hydrodynamic equations are solved using 
a so-called high-resolution shock-capturing scheme \cite{Font,Font2002,S2002}
in cylindrical coordinates (or on $y=0$ plane in Cartesian coordinates). 

The fundamental variables for geometry are: 
\beqn
\alpha &&: {\rm lapse~ function}, \nonumber \\
\beta^k &&: {\rm shift~ vector}, \nonumber \\
\gamma_{ij} &&:{\rm metric~ in~ 3D~ spatial~ hypersurface},\nonumber \\ 
\gamma &&=e^{12\phi}={\rm det}(\gamma_{ij}), \nonumber \\
\tilde \gamma_{ij}&&=e^{-4\phi}\gamma_{ij}, \nonumber \\
K_{ij} &&:{\rm extrinsic~curvature}. 
\eeqn
We evolve $\tilde \gamma_{ij}$, $\phi$, 
$\tilde A_{ij} \equiv e^{-4\phi}(K_{ij}-\gamma_{ij} K_k^{~k})$, 
and the trace of the extrinsic curvature $K_k^{~k}$ 
together with three auxiliary functions 
$F_i\equiv \delta^{jk}\pa_{j} \tilde \gamma_{ik}$ with an unconstrained
free evolution code as done in \cite{shibata,gr3d,bina,SN,S2002}.

The Einstein equations are solved in the Cartesian coordinates. 
To impose axisymmetric boundary conditions, 
the Cartoon method is adopted \cite{alcu}: 
Assuming a reflection symmetry with 
respect to the equatorial plane, we perform simulations 
using a fixed uniform grid with the grid 
size $N \times 3 \times N$ for $(x, y, z)$ which covers 
a computational domain as $0 \leq x \leq L$, $0 \leq z \leq L$, and
$-\Delta \leq y \leq \Delta$. 
Here, $N$ and $L$ are constants and $\Delta = L/N$. 
For $y= \pm \Delta$, the axisymmetric boundary conditions are imposed
using data sets on the $y=0$ plane. 

The slicing and spatial gauge conditions adopted in this paper are
basically the same as those in~\cite{shibata,gr3d,SBS,bina}; i.e., we
impose an ``approximately'' maximal slice condition ($K_k^{~k} \approx 0$)
and an ``approximately'' minimal distortion gauge condition [$\tilde D_i
(\pa_t \tilde \gamma^{ij}) \approx 0$ where $\tilde D_i$ is the
covariant derivative with respect to $\tilde \gamma_{ij}$] 
\cite{shibata,gr3d,bina}. 
In the approximately minimal distortion gauge condition, 
$F_i$ is zero in the linear order in
$h_{ij}(\equiv \tilde \gamma_{ij}-\delta_{ij})$
if $F_i=0$ initially. Thus, in the wave zone, 
$h_{ij}$ approximately satisfies a transverse condition $h_{ij,j}=0$.

We also performed a few simulations using a dynamical
gauge condition \cite{S03} in which we solve 
\beq
\pa_t \beta^k = \tilde \gamma^{kl} (F_l +\Delta t \pa_t F_l),
\label{dyn}
\eeq
where $\Delta t$ denotes a timestep in a numerical computation. 
We have found that the magnitude of the numerical error depends weakly on  
the spatial gauge condition, but gravitational
waveforms and qualitative nature of the numerical error do not. 
Thus, in this paper, we present the results obtained in the
approximately minimal distortion gauge condition. 

During numerical simulations, 
violations of the Hamiltonian constraint and conservation of
mass and angular momentum are monitored as code checks.
Numerical results for several test
calculations, including stability and collapse of nonrotating
and rotating neutron stars, have been described in \cite{S2002}.
Several convergence tests have been also presented in \cite{S2002}.

\subsection{Outer boundary conditions}

Outer boundary conditions for geometric variables
have been modified from previous ones \cite{shibata,gr3d,SBS,bina}. 
We impose a boundary condition for $\phi$ as 
\beqn
\phi ={M \over 2r} + O(r^{-2}), 
\eeqn
where $M$ denotes the ADM mass of a system computed at $t=0$. 
We note that $M$ is an approximately conserved quantity in axial symmetry, 
since only a small amount of gravitational waves are emitted
in axisymmetric oscillations. 

For $h_{ij}(=\tilde \gamma_{ij}-\delta_{ij})$, we first carry out
a coordinate transformation from the
Cartesian coordinates to spherical polar coordinates
$(r, \theta, \varphi)$ with the
flat metric, $\eta_{ab}$ 
[the subscripts $a$ and $b$ denote components of 
the spherical polar coordinates ($r, \theta, \varphi$)],
and then impose outgoing-wave outer boundary conditions of the form 
\beqn
&&h_{\hat r \hat r}r^3=f_1(t-r),\nonumber \\
&&h_{\hat r\hat \theta}r^2=f_2(t-r),\nonumber \\
&&h_{\hat r\hat \varphi}r^2=f_3(t-r),\nonumber \\
&&h_{\hat \theta \hat \theta}r=f_4(t-r),\nonumber \\
&&h_{\hat \theta \hat \varphi}r=f_5(t-r),\nonumber \\
&&h_{\hat \varphi \hat \varphi}r=f_6(t-r), \label{bc4}
\eeqn
where $h_{\hat a \hat b}$ denotes a tetrad component and $f_i~(i=1\sim 6)$ 
denote functions: Since $f_i(t-r)=f_i[t-\Delta t - (r-\Delta t)]$, 
the value of $f_i(t-r)$ at the outer boundaries 
should be equal to that at a time $t-\Delta t$ and 
at a radius $r-\Delta t$ which is determined 
using the values of 8 nearby grid points at a previous timestep \cite{SN}. 
These sets of the outer boundary conditions are well-suited for 
a solution of the linearized Einstein equation 
in the transverse-traceless gauge condition for $h_{ab}$
if they are imposed in a distant wave zone \cite{T82}. 
In the local wave zone, however, equations (\ref{bc4}) are
approximate boundary conditions since higher order terms of
$1/r$ are neglected. Therefore, 
numerical errors and unphysical solutions 
may be generated around the outer boundaries. 
In addition, $h_{ij}$ should physically contain nonwave modes
(such as stationary multipole modes) which do 
not obey the boundary conditions (\ref{bc4}). Since no
attention is paid to such modes in imposing the condition (\ref{bc4}), 
additional numerical errors may be excited. 

In the case of the approximately minimal distortion gauge, 
we adopt the same boundary conditions
for $F_i$, $K$, and $\tilde A_{ij}$ as 
those used in previous papers \cite{shibata,gr3d}:
i.e., $F_i=K=0$, and an outgoing-wave 
boundary condition is imposed for $\tilde A_{ij}$.
In the dynamical gauge condition, an outgoing-wave 
boundary condition is imposed for $F_i$, 
because it obeys a hyperbolic-type equation. 

\section{Wave extraction methods}

\subsection{Gauge-invariant technique}

Gravitational waves are extracted in terms of a gauge-invariant 
technique \cite{moncrief,Abrahams}. In this method, the fully nonlinear
3-metric $\gamma_{ab}$ in spherical polar coordinates
is split as $\eta_{ab}+\xi_{ab}$, where 
$\xi_{ab}$ is regarded as the perturbation on the flat background.
In axial symmetry, $\xi_{ab}$ can be decomposed  
as $\sum_{l} \zeta_{ab}^{l}$, where $\zeta_{ab}^{l}$ is given by 
\beqn
\zeta_{ab}^{l}=&& \left(
\begin{array}{lll}
\displaystyle 
H_{2l} Y_{l0} & h_{1l} Y_{l0,\theta}& 0\\
\ast& r^2(K_{l}Y_{l0}+G_{l}W_{l0}) & 0 \\
\ast& \ast&r^2\sin^2\theta(K_{l}Y_{l0}-G_{l}W_{l0}) \\
\end{array}
\right) \nonumber \\
&&+\left(\begin{array}{ccc}
0 &  0 & C_{l} \pa_{\theta}Y_{l0}\sin\theta  \\
\ast & 0 & -r^2 D_{l}W_{l0}\sin\theta  \\
\ast & \ast & 0 \\
\end{array}
\right). \label{eqpert}
\eeqn
Here, $\ast$ denotes the symmetric components. The first term
in Eq. (\ref{eqpert}) corresponds to even parity (polar)
perturbations and the second one to odd parity (axial) perturbations. 
The quantities $H_{2l}$, $h_{1l}$, $K_{l}$, $G_{l}$, $C_{l}$, and 
$D_{l}$ are functions of $r$ and $t$, and are calculated 
by performing integrations over a two-sphere of a given radius.
$Y_{l0}$ is the spherical harmonic function, and 
$W_{l0}$ is defined as 
\beqn
W_{l0} \equiv \Bigl[ (\pa_{\theta})^2-\cot\theta \pa_{\theta}
\Bigl] Y_{l0}. 
\eeqn

The gauge-invariant variables of even and odd parities are then 
defined as \cite{moncrief,Abrahams}
\beqn
&&R_{l}^{\rm E}(t,r) \equiv 
\sqrt{2(l-2)! \over (l+2)!}
\Bigl\{ 2 k_{2l}+l(l+1)k_{1l} \Bigr\}, \\
&&R_{l}^{\rm O}(t,r) \equiv \sqrt{2(l+2)! \over (l-2)!}
\biggl({C_{l} \over r}+r \pa_r D_{l}\biggr),
\eeqn
where
\beqn
k_{1l}&& \equiv K_{l}+l(l+1)G_{l}+2r \pa_r G_{l}-2{h_{1l} \over r},\\
k_{2l}&& \equiv H_{2l}  - {\pa \over \pa r}
\Bigl[r\{ K_{l}+l(l+1)G_{l} \} \Bigr].
\eeqn
Luminosity of gravitational waves is computed from
\beqn
{dE \over dt}={r^2 \over 32\pi}\sum_{l}\Bigl[
|\pa_t R_{l}^{\rm E}|^2+|\pa_t R_{l}^{\rm O}|^2 \Bigr]. 
\label{dedt} 
\eeqn
The time derivative of the gauge-invariant quantities in Eq. (\ref{dedt}) 
is taken by a finite-differencing. 
Hereafter, we focus only on the even-parity mode with $l=2$ 
because for the oscillations considered in this paper, 
its amplitude is much larger than that of other modes. 

In an appropriate radiation gauge,
$+$-mode of gravitational waves is extracted 
from the asymptotic behavior of the following quantity
at $r\rightarrow \infty$; 
\beqn
h_+ \equiv {1 \over 2r^2}\biggl(\gamma_{\theta\theta}-
{\gamma_{\varphi\varphi} \over \sin^2\theta}\biggr). 
\eeqn
For $l=2$, $h_+$ can be written as
\beqn
h_+={A_2(t) \over r}\sin^2\theta, 
\eeqn
where $A_2(t)$ denotes a function of time. Using $A_2$, 
the asymptotic behavior of $R_2^{\rm E}$ is written as
\beq
R_2^{\rm E} \rightarrow \sqrt{{64\pi \over 15}}{A_2 \over r}. 
\eeq

\subsection{Quadrupole formula}

In quadrupole formulas, gravitational waves are computed from 
\beq
h_{ij}=P_{i}^{~k} P_{j}^{~l} \biggl(
{2 \over r}{d^2\bI_{kl} \over dt^2}\biggr),\label{quadf}
\eeq
where $\bI_{ij}$ and $P_i^{~j}=\delta_{ij}-n_i n_j$ ($n_i=x^i/r$)
denote a tracefree quadrupole moment and a projection tensor. 
From this expression, $+$-mode of gravitational waves with $l=2$
in axial symmetry is written as
\beq
h_+^{\rm quad} = {\ddot I_{xx}(t_{\rm ret}) - \ddot I_{zz}(t_{\rm ret})
\over r}\sin^2\theta, \label{quadr}
\eeq
where $I_{ij}$ denotes an appropriately defined
quadrupole moment, and $\ddot I_{ij}$ its second time derivative.
Equation (\ref{quadr}) implies that in quadrupole formulas, 
$A_2(t)=\ddot I_{xx}(t_{\rm ret}) - \ddot I_{zz}(t_{\rm ret})$. 

$t_{\rm ret}$ is a retarded time which is approximately defined by 
\beq
t_{\rm ret}=t - r_{\rm circ} - 2M \ln \Big({r_{\rm circ} \over 2M}-1\Big),
\label{tret}
\eeq
where $r_{\rm circ}= r(1+M/2r)^2 \gg M$. 
Equation (\ref{tret}) is the valid expression only for the Schwarzschild
spacetime. In axisymmetric spacetimes, the retarded time
should depend on the direction of wave propagation. 
However, the magnitude of the difference between
Eq. (\ref{tret}) and the exact one would be of $O(M)$ and, hence,
much smaller than the typical wave length. 

In fully general relativistic and dynamical spacetimes, 
there is no unique definition for the quadrupole moment and, 
hence, for $\ddot I_{ij}$. Here, we choose for simplicity 
\beq
I_{ij} = \int \rho_* x^i x^j d^3x. 
\eeq
Then, using the continuity equation of the form 
\beq
\pa_t \rho_* + \pa_i (\rho_* v^i)=0, 
\eeq
we can compute the first time derivative as 
\beq
\dot I_{ij} = \int \rho_* (v^i x^j +x^i v^j)d^3x.
\eeq
To compute $\ddot I_{ij}$, we carried out a finite differencing
of the numerical result for $\dot I_{ij}$. 

Since the quadrupole moment $I_{ij}$ is not defined 
uniquely in the dynamical spacetime in general relativity, 
gravitational waveforms computed by quadrupole formulas
depend on the form of $I_{ij}$. In addition, they depend on 
the gauge conditions, since a physical point 
which coordinates $x^i$ and $t$ denote 
(and, as a result, the magnitude of $I_{ij}$ and $\ddot I_{ij}$) 
is not identical in two different gauge conditions. 
Even if an identical definition of the quadrupole formula is employed,
waveforms do not in general agree when different gauge conditions are 
adopted. Therefore, we should keep in mind that 
the waveforms computed from Eq. (\ref{quadr}) 
are special ones obtained in specific choices of $I_{ij}$ and 
the gauge condition. All these facts imply that to know 
the validity of the quadrupole formula which one chooses,
comparison between the waveforms by the quadrupole formula
with those extracted from metric should be made. 

\section{Numerical Results}

\subsection{Setting}

The simulations were carried out along the following procedure: 
(1) neutron stars in equilibrium were prepared, 
(2) a perturbation to the equilibria was added, 
(3) the constraint conditions were re-imposed by solving the 
constraint equations for the perturbed state, 
and then (4) we started the time evolution. 

To model neutron stars in equilibrium, we simply adopt the 
polytropic equation of state as 
\beq
P=K\rho^{1+{1 \over n}}, 
\eeq
where $K$ is the polytropic constant and $n$ the polytropic index. 
For the evolution of the neutron stars, 
we adopt a $\Gamma$-law equation of state in the form 
\beq
P=(\Gamma-1)\rho \varep,
\eeq 
where $\Gamma=1+1/n$. 
We set $n=1~(\Gamma=2)$ as a reasonable qualitative approximation
to cold, nuclear equation of state.
%%%%%%%% new 
With these equations of state, realistic neutron stars may not
be modeled precisely. However, in the current situation that 
no one know a real equation of state of neutron stars exactly, 
modeling neutron stars with a simple equation of state is an
adequate and popular strategy. 

In the present choice of the equation of state, 
physical units enter the problem only through the polytropic constant $K$,
which can be chosen arbitrarily or else completely scaled out of 
the problem.  For $n=1$, $K^{1/2}$ has units of length, time, and mass, 
and $K^{-1}$ units of density in the geometrical units. Using this
property, we rescale all the quantities to be nondimensional 
and show only the nondimensional quantities. 

One often prefers to use particular dimensional units even in
the polytropic equation of state. 
For example, in \cite{Font2002}, the authors fix the value of $K$ as
$1.455 \times 10^{5}$ cgs. 
For the sake of comparison with the previous paper, 
we convert non-dimensional quantities to dimensional 
ones with the polytropic constant $K = 1.455 \times 10^{5}$ cgs. 
In this special value, the mass, the density, and the time in the 
dimensional units are written as 
\beqn
&& M_{\rm dim}=1.80 M_{\odot}
\biggr({K \over 1.455\times 10^5~{\rm cgs}}\biggr)^{1/2}
\biggl({M \over 0.180}\biggr),\\
&& \rho_{\rm dim}=1.86 \times 10^{15}~{\rm g/cm^3} 
\biggr({K \over 1.455\times 10^5~{\rm cgs}}\biggr)^{-1}
\biggl({\rho \over 0.300}\biggr),\\
&& T_{\rm dim}=4.93~{\rm msec} 
\biggr({K \over 1.455\times 10^5~{\rm cgs}}\biggr)^{1/2}
\biggl({T \over 100}\biggr). 
\eeqn

We adopted six models of neutron stars referred to as S1--S3 and R1--R3
in this numerical experiment. Models S1--S3 are spherical stars and 
R1--R3 are rigidly and rapidly rotating stars 
whose angular velocities at the equatorial surface are approximately
equal to the Keplerian angular velocity (i.e., at mass-shedding limits). 
By exploring rotating stars at the mass-shedding limits, 
the effects of rotation on rigidly rotating neutron stars are
clarified most efficiently. 

The maximum gravitational masses of spherical stars and rigidly rotating 
stars with $n=1~(\Gamma=2)$ are $\approx 0.164K^{1/2}$ and 
$\approx 0.188K^{1/2}$, respectively \cite{CST}. 
Thus, the models adopted here are sufficiently general relativistic 
in the sense that their masses are close to the maximum values. 
In Table I, characteristic quantities for models S1--S3 and R1--R3 
are listed in the nondimensional units
(in the units with $c=G=K=1$). 

\begin{table}[t]
\begin{center}
\begin{tabular}{|c|c|c|c|c|c|c|} \hline
& $\rho_c$ & $M_*$ & $M$ & $M/R$ & $J/M^2$ &
$P_{\rm osc}/(2\pi R^{3/2}M^{-1/2})$ \\ \hline
S1 & 0.127 & 0.150 & 0.140 & 0.146 & 0    & 0.84 \\ \hline
S2 & 0.191 & 0.170 & 0.156 & 0.178 & 0    & 0.84\\ \hline
S3 & 0.255 & 0.178 & 0.162 & 0.200 & 0    & 0.85\\ \hline
R1 & 0.103 & 0.169 & 0.158 & 0.111(0.181) & 0.667 & 0.48 \\ \hline
R2 & 0.118 & 0.178 & 0.165 & 0.120(0.194) & 0.648 & 0.48 \\ \hline
R3 & 0.136 & 0.186 & 0.172 & 0.129(0.207) & 0.630 & 0.50 \\ \hline
\end{tabular}
\caption{Central density $\rho_c$, baryon rest-mass $M_*$, 
ADM mass $M$, compactness $M/R$, angular momentum $J$ in units of $M^2$, 
and numerical results for fundamental radial oscillation period 
with $l=2$ in units of $2\pi R^{3/2}/M^{1/2}$ of 
neutron stars that we pick up in this paper.
Here, $R$ denotes the circumference radius at the equatorial surface. 
For the rotating stars, the compactness measured by the polar
radius is also listed in the bracket. 
All the quantities are shown in units of $c=G=K=1$.
}
\end{center}
\vspace{-5mm}
\end{table}

To induce nonspherical stellar oscillations
to nonrotating stars, we superimposed a velocity perturbation as 
\beq
\delta u_{\varpi} = -V \varpi~~~{\rm and}~~~\delta u_z = V z,
\eeq
where $\delta u_{\varpi}$ and $\delta u_z$ denote the 
four-velocity of cylindrical ($\varpi=\sqrt{x^2+y^2}$) and $z$ components.  
$V$ is a constant and put as $V=0.1/\varpi_e$ where 
$\varpi_e$ denotes the coordinate radius at the equator: i.e.,
at the equatorial surface, the velocity is $\approx 10\%$ of the 
light speed. 

For the rotating stars, two types of the perturbations are adopted. 
One is a velocity perturbation given by 
\beq
\delta u_{\varpi} = - {V \over 2} \varpi~~~{\rm and}~~~\delta u_z = V z,
\eeq
with $V=0.3/\varpi_e$. The other is a pressure perturbation in which 
we simply reduced the pressure uniformly by changing the polytropic 
constant from the equilibrium value to a smaller one.
In this case, a quasiradial oscillation is induced. 
Since the rotating star is nonspherical, gravitational waves are 
emitted even in this setting. 

The simulations were performed changing grid spacing and location 
of outer boundaries. As found in \cite{S2002}, the numerical
results are sufficiently convergent if the stellar radius is covered by 
more than 30 grid points. Taking into account this fact, 
the grid spacing is fixed in the simulations of this paper as follows: 
For nonrotating stars, we chose it as $\varpi_e/40$,
and for rotating stars, $\varpi_e/60$: Since the 
axial ratio of the rotating stars at the mass-shedding
limit is $\approx 0.59$, the polar axis is initially
covered by about 36 grid points with this setting. 

On the other hand, the location of the outer boundaries was changed 
varying $N$ from 480 to 720. 
We typically choose $N=720$, since with this number, $L$ is larger than 
a characteristic gravitational wave length $\lambda$ and, thus, 
the outer boundaries are located in a local wave zone. 
Since $L > \lambda$, we expect that the amplitude of
gravitational waves could be calculated within $\sim 10$\% error. 

Table II describes the values of $L$ and the location $L_e$ at which
the gauge-invariant variables are extracted. 
Typically, $L_e$ is chosen to be $\sim 0.9L$. Note that 
varying $L_e$ from $L/2$ to $0.95L$, it is found that 
the amplitude of gravitational waves depends very weakly 
on the location of $L_e$ [see Figs. 1(b) and 2(b)]
for a fixed value of $L$. 

\begin{table}[t]
\begin{center}
\begin{tabular}{|c|c|c|c|} \hline
Nonrotational & $L/M$ for $N=480$, 600, 720& $L_e/M$ for $N=480$, 600, 720
& $\lambda/M$ \\ \hline
S1 & 64.4,~~80.5,~~96.6  & 53.5,~~66.9,~~80.3  & 94.8 \\ \hline
S2 & 57.5,~~71.9,~~86.3  & 47.8,~~59.8,~~71.8  & 70.5 \\ \hline
S3 & 55.4,~~69.3,~~83.2  & 46.1,~~57.6,~~69.2  & 60.5 \\ \hline
\end{tabular}
\begin{tabular}{|c|c|c|c|} \hline
Rotational & $L/M$ for $N=480$, 720& $L_e/M$ for $N=480$, 720
& $\lambda/M$ \\ \hline
R1 & 63.3,~~94.9  & 57.9,~~86.9    & 80.8 \\ \hline
R2 & 58.0,~~87.0  & 53.1,~~79.6    & 73.3 \\ \hline
R3 & 53.2,~~79.8  & 48.7,~~73.0    & 66.5 \\ \hline
\end{tabular}
\caption{Values of $L$ and $L_e$ in units of $M$ for various values of $N$. 
For comparison, we show a gravitational wave length for 
the fundamental quadrupole ($l=2$) mode derived from numerical results.  
}
\end{center}
\vspace{-5mm}
\end{table}

\subsection{Systematic numerical errors}

Since the outer boundaries are located 
in the local wave zone and, hence, the boundary conditions
which are appropriate only for the distant wave zone are 
not precise ones, systematic numerical errors are generated around 
the outer boundaries. As a consequence,
gravitational waveforms (gauge-invariant variables)  
are contaminated by numerical errors. To accurately 
extract gravitational waves, we need to eliminate the errors 
from raw data sets of the gauge-invariant variables. 

\begin{figure}[htb]
\vspace*{-4mm}
\begin{center}
\epsfxsize=2.9in
\leavevmode
(a)\epsffile{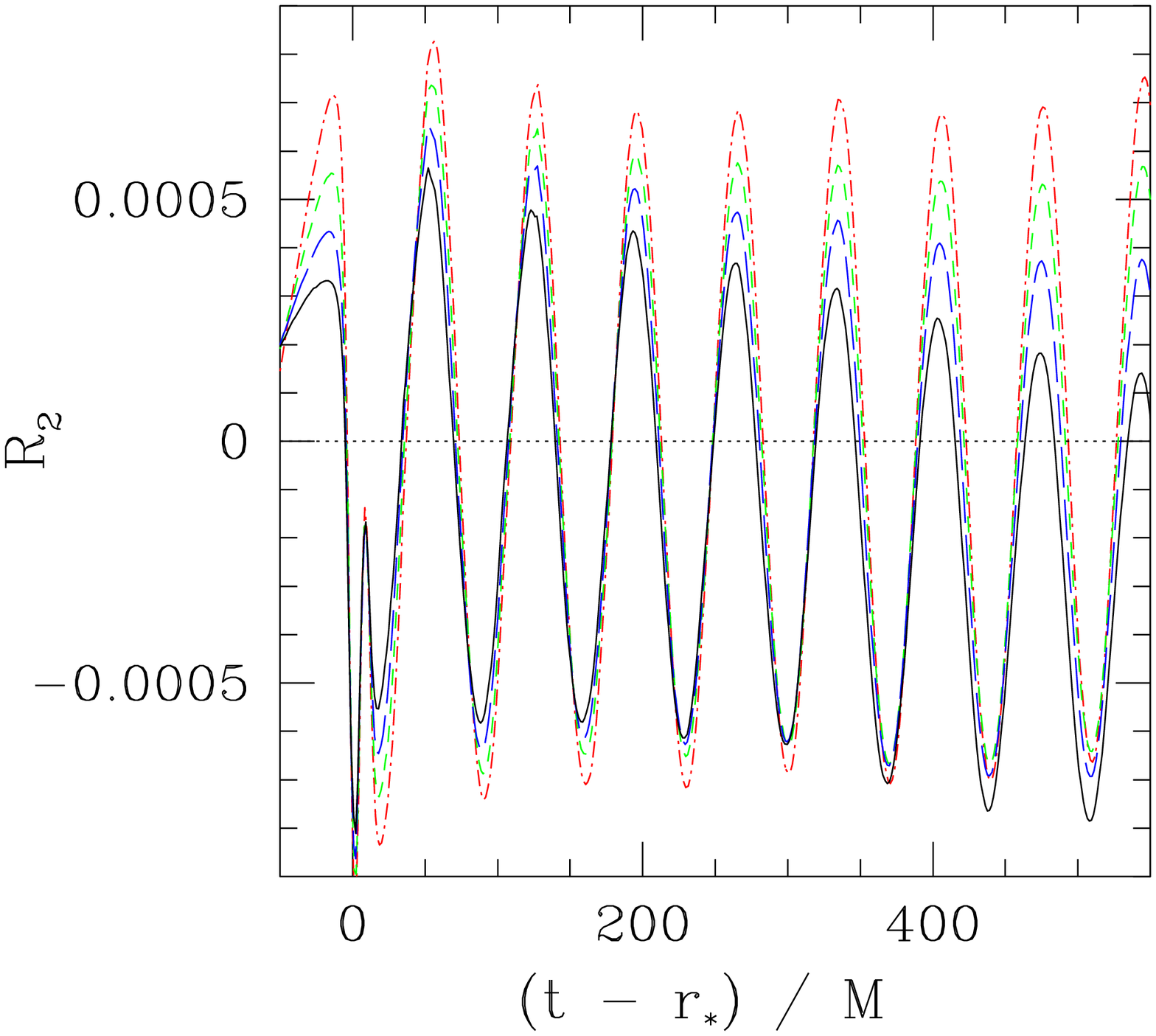}
\epsfxsize=2.9in
\leavevmode
~~~(b)\epsffile{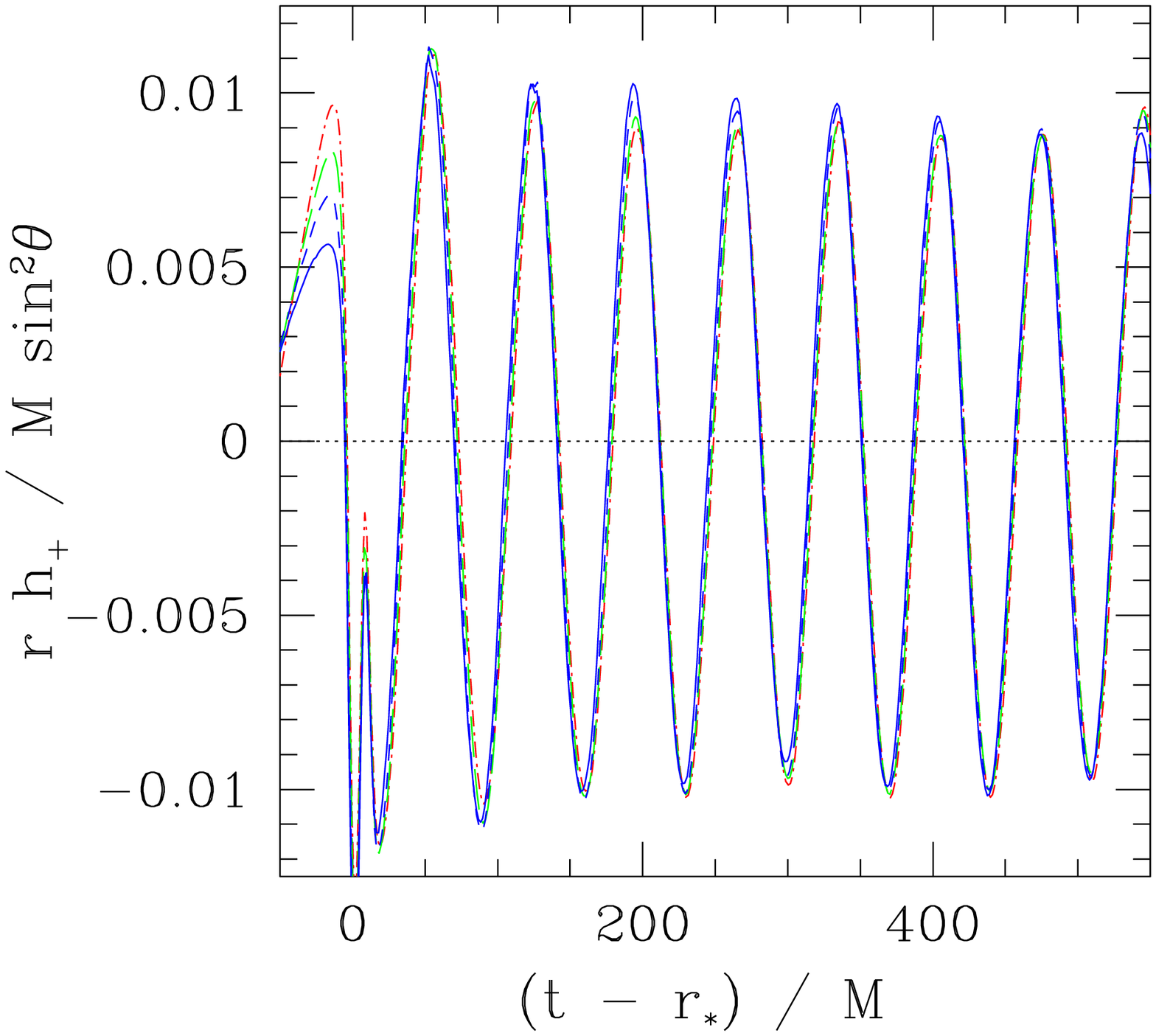}
%\vspace*{-4mm}
\caption{(a) Gauge-invariant variables with $l=2$ 
in units of $M$ for oscillation of a nonrotating neutron star S2 
extracted at $r/M=50.2$ (dotted-dashed curve), 
57.4 (dashed curve), 64.6 (long-dashed curve), and 71.8 (solid curve).
(b) gravitational waveforms after systematic numerical errors 
are subtracted. In the simulation, $N=720$. 
\label{FIG1}
}
\end{center}
\end{figure}

\begin{figure}[htb]
\vspace*{-4mm}
\begin{center}
\epsfxsize=2.9in
\leavevmode
(a)\epsffile{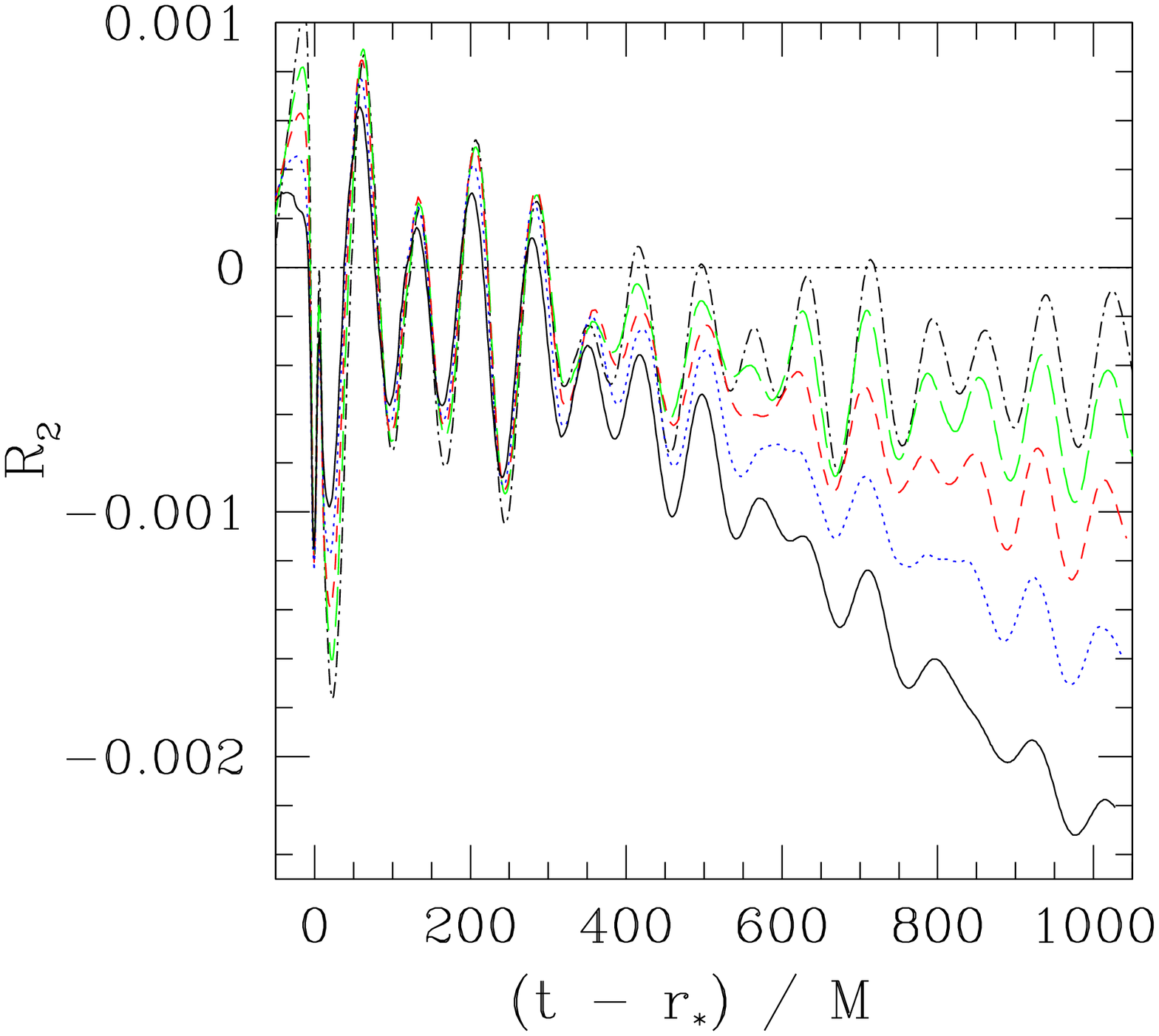}
\epsfxsize=2.9in
\leavevmode
~~~(b)\epsffile{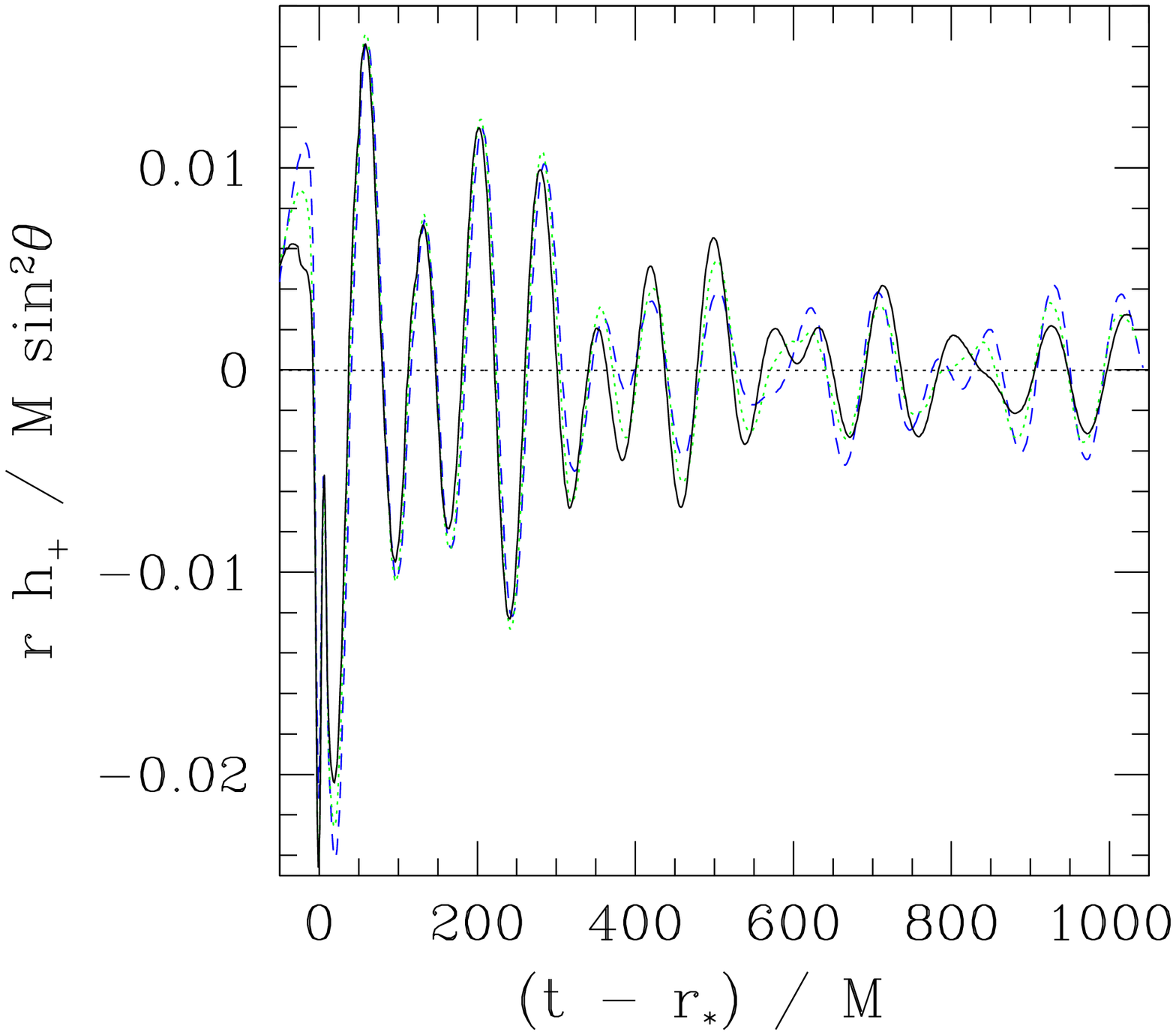}
%\vspace*{-4mm}
\caption{(a) Gauge-invariant variables with $l=2$ in units of $M$ for
oscillation of a rotating neutron star R2 extracted
at $r/M=50.7$ (dotted-dashed curve), 57.9 (long-dashed curve), 
65.2 (dashed curve), 72.4 (dotted curve), and 79.6 (solid curve). 
(b) gravitational waveforms after systematic numerical errors are 
subtracted at $r/M=$65.2 (dashed curve), 72.4 (dotted curve), and 
79.6 (solid curve). The simulation was performed with $N=720$. 
\label{FIG2}
}
\end{center}
\end{figure}

First, we summarize the behavior of the numerical errors. 
In Figs. 1(a) and 2(a), we display time evolution of 
raw data sets of the gauge-invariant variables extracted at 
several radii for models S2 and R2 with $N=720$. 
The gauge-invariant variables are composed mainly of three components; 
(i) a pure wave component which denotes gravitational waves, 
(ii) a constant component, and (iii) a secular drift component
whose magnitude increases with time and is larger for the larger value
of radius. To obtain clear gravitational waveforms, it is necessary to
eliminate the components (ii) and (iii). 

The presence of the component (ii) is in part due to the fact that 
the gauge-invariant variables are computed at finite radii.
They are composed not only of gravitational waves but also of
a quasistationary component which falls off as $r^{-n}$ for $n \geq 2$.
In particular, rapidly rotating stars are of spheroidal 
shape and, hence, they yield quasistationary quadrupole and 
higher multipole moments which slowly vary throughout the simulation. 
As Fig. 2(a) shows (see for $t - r_* < 0$),
this component is smaller for the larger value of 
extracted radius because it falls off as $r^{-n}$ with $n \geq 2$. 

A part of the component (ii) and the component (iii) are 
numerical errors associated with an approximate treatment of the
outer boundary conditions.
In the following, we explain the origin of it in detail. 

In the wave zone, the magnitude of $h_{ij}$ is small and, hence, 
it approximately obeys a linearized 
Einstein equation. In the formalism which we currently use \cite{bina},
the linearized equation for $h_{ij}$ is of the form 
\beqn
\ddot h_{ij}=\Delta_f h_{ij}+S_{ij}^h,
\eeqn
where $\Delta_f$ denotes the flat Laplacian and $S_{ij}^h$ is
composed of spatial derivatives of $a \equiv \alpha-1$, 
$\dot \beta^i$, $p \equiv e^\phi-1$, and $F_k$ as 
\beqn
S_{ij}^h=&&(4p+2a)_{|ij}-{1 \over 3} \eta_{ij}\Delta_f (4p+2a) \nonumber \\
&&+\eta_{ik}\dot \beta^k_{~|j}+\eta_{jk}\dot \beta^k_{~|i}
-{2 \over 3} \eta_{ij} \dot \beta^k_{~|k} 
-\Big(F_{i|j}+F_{j|i}-{2 \over 3} \eta_{ij} \eta^{kl}F_{k|l}\Big). 
\eeqn
Here, $|i$ denotes the covariant derivative with respect to the
flat metric $\eta_{jk}$. 

Because of the functional form of $S_{ij}^h$, 
the solution of $h_{ij}$ may be written as
\beq
h_{ij}=h_{ij}^{\rm GW}
+ \xi_{i|j} + \xi_{j|i} -{2 \over 3}\eta_{ij}\xi^k_{~|k},
\eeq
where $h_{ij}^{\rm GW}$ and $\xi_k$ obey the following equations: 
\beqn
&&\ddot h_{ij}^{\rm GW}=\Delta_f h_{ij}^{\rm GW},\label{linear} \\
&&\ddot \xi_k=\Delta_f \xi_k +\dot \beta_k + (2p+a)_{|k}-F_k. \label{xieq} 
\eeqn
Since the inhomogeneous solution of $h_{ij}$, which is 
associated with $S_{ij}^h$, is written by a gauge variable $\xi_k$,
$S_{ij}^h$ does not contribute the gauge-invariant variable\footnote{The gauge-invariant variables are
computed from $\gamma_{ij}-\eta_{ij} \equiv H_{ij}$.
In the linear order, $H_{ij} = h_{ij}+4 p\eta_{ij}$. 
The linearized Einstein equation for $H_{ij}$
in our formulation \cite{shibata,bina}
is 
\beqn
\ddot H_{ij} &&= \Delta_f H_{ij} + (4p+2a)_{|ij} 
+\eta_{ik}\dot \beta^k_{~|j}+\eta_{jk}\dot \beta^k_{~|i}
-\Big(F_{i|j}+F_{j|i}\Big)
-{2 \over 3}\eta_{ij}(8\Delta_f p - \eta^{kl}F_{k|l}). 
\eeqn
If the Hamiltonian constraint in the linear level is satisfied,
$8\Delta_f p - \eta^{kl}F_{k|l}=0$. Then,
the solution of $H_{ij}$ is $\bar h_{ij} + \xi_{i|j} + \xi_{j|i}$,
and consequently, the gauge-invariant variables do not
contain $\xi_i$. However, 
if the Hamiltonian constraint is violated, 
the effects of $\xi_i$ is contained in the gauge-invariant variables. 
In the present work, we found that its effect is small, and
hence we ignore it. }. 

Gravitational waves may be extracted from $G_2$, which is calculated by 
\beq
G_2={1 \over 48}
\oint (h_{\hat \theta \hat \theta}- h_{\hat \varphi \hat \varphi})
W_{20} dS. 
\eeq
This variable can be regarded as gravitational waves in gauge conditions 
in which $\xi_k=0$ (e.g., in the harmonic gauge condition).  
However, in the present case, $\xi_k$ is not guaranteed to be
vanishing and the effects of $\xi_k$ are contained in $G_2$. 
Consequently, the waveforms may be deformed by unphysical modulations
and secular drifts due to the presence of $\xi^k$. 
This illustrates that the gauge-invariant technique plays an 
important role for extraction of gravitational waves 
in general gauge conditions. 

General forms of outgoing-wave solutions of Eq. (\ref{linear})
for $h_{ij}^{\rm GW}$ 
have been derived by Teukolsky \cite{T82} and Nakamura \cite{N84}.
However, numerically, such solutions can be exactly computed only when
(A) a strict outgoing-wave boundary condition well-suited
in a local wave zone is imposed and 
(B) the transverse condition is guaranteed.
In the present numerical simulations, 
these conditions are not satisfied strictly. Therefore, unphysical modes
as well as numerical errors contaminate 
numerical solutions of $h_{ij}^{\rm GW}$. 

One of the candidates for the dominant unphysical modes 
is a solution for the equations $\Delta_f h_{ij}^{\rm GW}=0$ and 
$\ddot h_{ij}^{\rm GW}=0$. From the relation
$\ddot h_{ij}^{\rm GW}=0$, $h_{ij}^{\rm GW}$ is written as
$H_{ij}^0(x)+H_{ij}^1(x)t$ where 
$H_{ij}^{0}$ and $H_{ij}^{1}$ satisfy the Laplace equation. 
Note that $H^n_{ij}~(n=0$ and 1) do not 
satisfy the transverse-traceless condition in general. 
By performing a spherical harmonic decomposition of $h_{ij}$ in the 
spherical polar coordinates (see Appendix A), 
we find the asymptotic behavior of 
the solutions of $l$-th moment for $r \gg M$ as 
\beqn
H_{ij}^n \rightarrow r^{l \pm 2}, r^l, r^{-l \pm 1},
~{\rm and}~r^{-l-3}. \label{hijeq}
\eeqn
The solutions of $r^{l-2}$ and $r^{-l-3}$
can be written in the form $V_{i|j}+V_{j|i}$ where $V_i$ denotes a vector.
These components are eliminated from the gauge-invariant variables. 
However, other solutions cannot be written in terms of $V_i$.
Thus, four of six solutions may 
be contained in the gauge-invariant variables as unphysical modes. 
Properties of the unphysical solutions are summarized as follows:
(a) they may increase linearly with time and
(b) they may be larger for larger extracted radii because 
of the presence of the modes proportional to $r^{l+2}$ and $r^l$. These 
properties agree with the numerical results shown in Figs. 1(a) and 2(a). 

Besides the global numerical errors described above, 
unphysical local waves 
of the form $f(t \pm r)$ may be contained in $h_{ij}^{\rm GW}$.
However, this is not a systematic error and, hence, 
it is not possible to eliminate systematically. 
Fortunately, the magnitude of such components is not as large as that of 
the systematic errors (see below). 

The systematic non-wave components may be 
fitted using a function of the form $C = C_0(r) + C_1(r) t$.
Thus, we determine $C_0$ and $C_1$ 
and then subtract $C$ from the gauge-invariant variables. 
As mentioned above, $C_0(r)$ arises both from a 
nonwave component associated with the stationary quadrupole moment
and from the unphysical modes associated with $H_{ij}^n$. 
$C_1(r) t$ arises from the unphysical modes associated with $H_{ij}^n$. 

For the nonrotating stars, $C_0$ and $C_1$ appear to be unchanged
throughout the simulations. Thus, 
to determine $C_0(r)$ and $C_1(r)$ at each radius, a least-square 
fitting against the gauge-invariant variables is 
carried out in the time domain. 
For the fitting, all the data sets with $t-r_* \geq 0$ are used. 

For the rotating stars, on the other hand, $C_0$ and, in particular, $C_1$ 
suddenly change at $t-r_* \sim 300M$.
(The reason is not very clear.) Thus, these coefficients are
separately determined for $t-r_* \agt 300M$ and $t-r_* \alt 300M$, 
carrying out the least-square fitting with two different data sets.
Namely, we subtract a function of the form 
\beqn
\left\{
\begin{array}{ll}
\displaystyle 
C_0(r)+C_1(r)t &~{\rm for}~t \leq t_m(r),\\
C'_0(r)+C'_1(r)t &~{\rm for}~t \geq t_m(r),
\end{array}
\right.
\eeqn
where $t_m(r)$ is a time $\sim 300M$ and satisfies the relation
$C_0(r)+C_1(r)t_m(r)=C'_0(r)+C'_1(r)t_m(r)$ at each radius. 

To validate that modified waveforms depend weakly on the
subtraction method, the following alternative method is also adopted: 
According to Eq. (\ref{hijeq}), $C_0$ and $C_1$ for $R_2^{\rm E}$ are
expressed by linear combinations of the functions
$r^4$, $r^2$, $r^{-1}$, and $r^{-3}$.
For a large value of the extraction radius $L_e$,
we may expect that $C_0 \approx 0$ and $C_1 \propto r^4$ 
at the leading order. In this assumption, $C$ may be computed as
\beq
C \approx
{r_2 R_2^{\rm E}(r_2)-r_1 R_2^{\rm E}(r_1) \over r_2^5-r_1^5} L_e^5, 
\eeq
where $r_1$ and $r_2$ denote two different radii which is close to $L_e$. 
After the subtraction of the dominant part, it is found that 
a component of small magnitude 
associated with $C_0$ still remains. To eliminate this remaining
nonwave part, we simply subtract a constant from the resulting 
waveform. 

In addition to the least-square fitting method, 
we have also adopted this method and confirmed that the subtracted
waveforms by this alternative method agree approximately with those 
in the least-square fitting (the wave phases agree well, 
and relative difference of the amplitude is within 10\%).
Thus, in the following, numerical results based on the 
subtraction using the least-square fitting are presented. 

In Figs. 1(b) and 2(b), we display improved 
gravitational waveforms $rh_+/\sin^2\theta$ 
obtained after the nonwave components and the numerical errors 
are subtracted. It is found that the resulting wave amplitude and phase
depend very weakly on the extracted radii for $t - r_* > 0$. This confirms
that the extracted components are certainly gravitational waves. 

For the rotating star, the amplitudes of 
the gravitational waveforms extracted at different radii
are in slight disagreement each other. 
Figure 2(b) shows that the magnitude of the
difference is $\sim 10^{-3}$. 
This indicates that even in the improved waveforms,
numerical errors of magnitude $\sim 10^{-3}$ still remain.
The origin is likely to be a nonsystematic error such as
spurious wave components. 

Besides quasiperiodic waves, a spike is visible at the beginning of
the simulation (at $t \sim 10M$) in Figs. 1 and 2. 
We do not understand the origin of it exactly. The following is 
inference for the possible origin: 
At t=0, we rather crudely add a weakly nonlinear velocity
perturbation to equilibrium states. 
Because of the weak nonlinearity, impulsive gravitational
waves may be excited besides eigen oscillation modes of the
equilibrium stars. Such impulsive gravitational waves seem to
propagate at $t - r \sim 0$. 

Before closing this subsection, we note the following: 
The magnitude of the numerical error associated with $H_{ij}^n$ 
is not very large in three-dimensional simulations 
for binary neutron star merger \cite{bina} and oscillation of 
neutron stars \cite{gr3d}, although it might be contained. 
We suspect that the excitation of such unphysical modes
may be associated with the interpolation used in the Cartoon method. 
(In this paper, we simply adopt a second-order interpolation.) 
This suggests that there would be still a room to improve the 
interpolation scheme in our numerical implementation.

\subsection{Gravitational waveforms}

\subsubsection{Nonrotating stars}

Improved gravitational waveforms with $l=2$ from 
axisymmetrically oscillating and nonrotating neutron stars 
for models S2 and S3 are displayed in Fig. 3. 
For both models, the simulations were 
performed with $N=480$ (dotted-dashed curves), 600 (long-dashed curves), 
and 720 (solid curves). Gravitational waveforms
evaluated by the quadrupole formula (dashed curves) are displayed
together to illustrate its validity. The quadrupole formula was used
in all the simulations and it is found that the gravitational
waveforms depend very weakly on the value of $N$. 
Here, the numerical results for $N=720$ are plotted. 

\begin{figure}[htb]
\vspace*{-4mm}
\begin{center}
\epsfxsize=2.9in
\leavevmode
(a)\epsffile{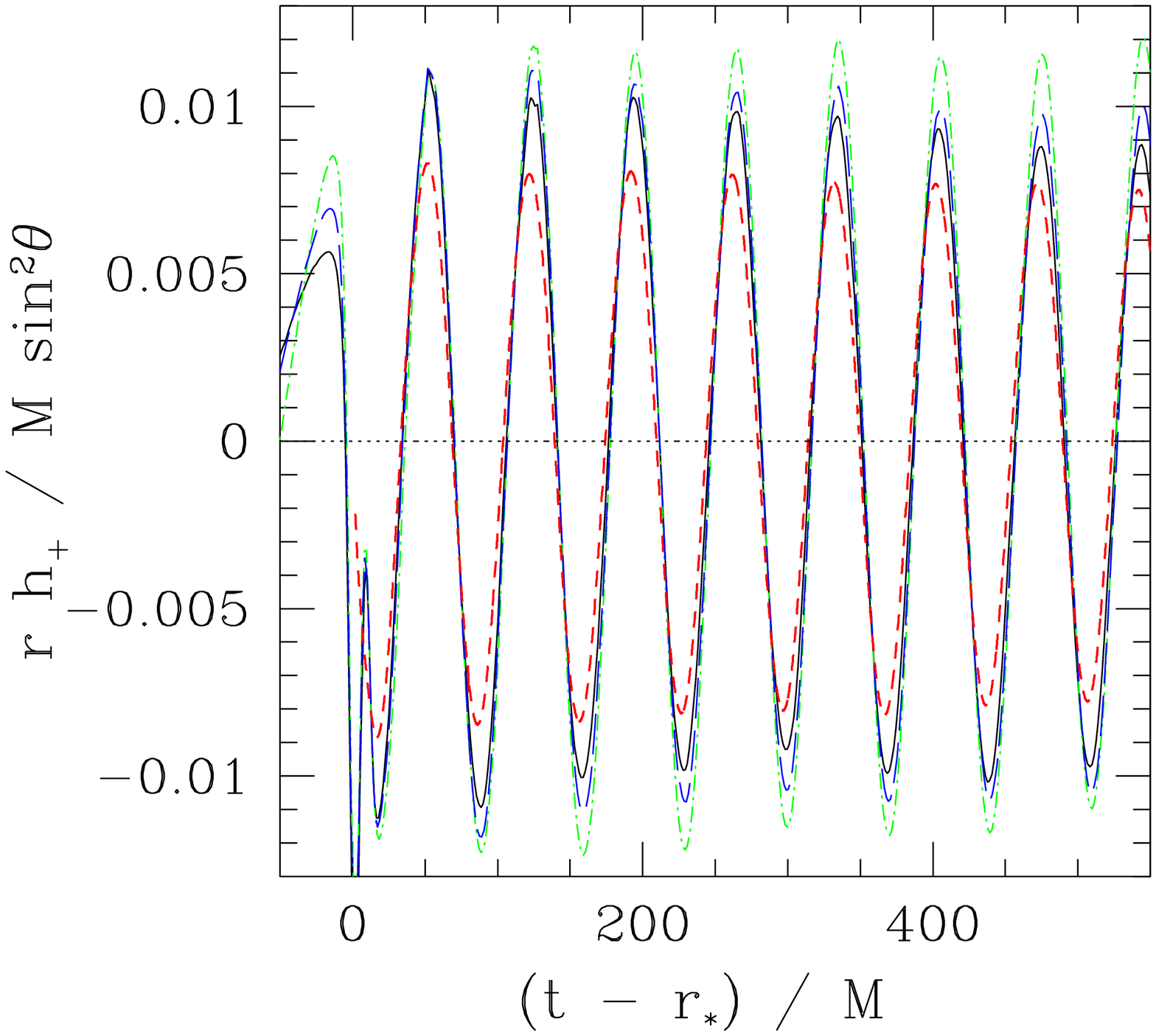}
\epsfxsize=2.9in
\leavevmode
~~~(b)\epsffile{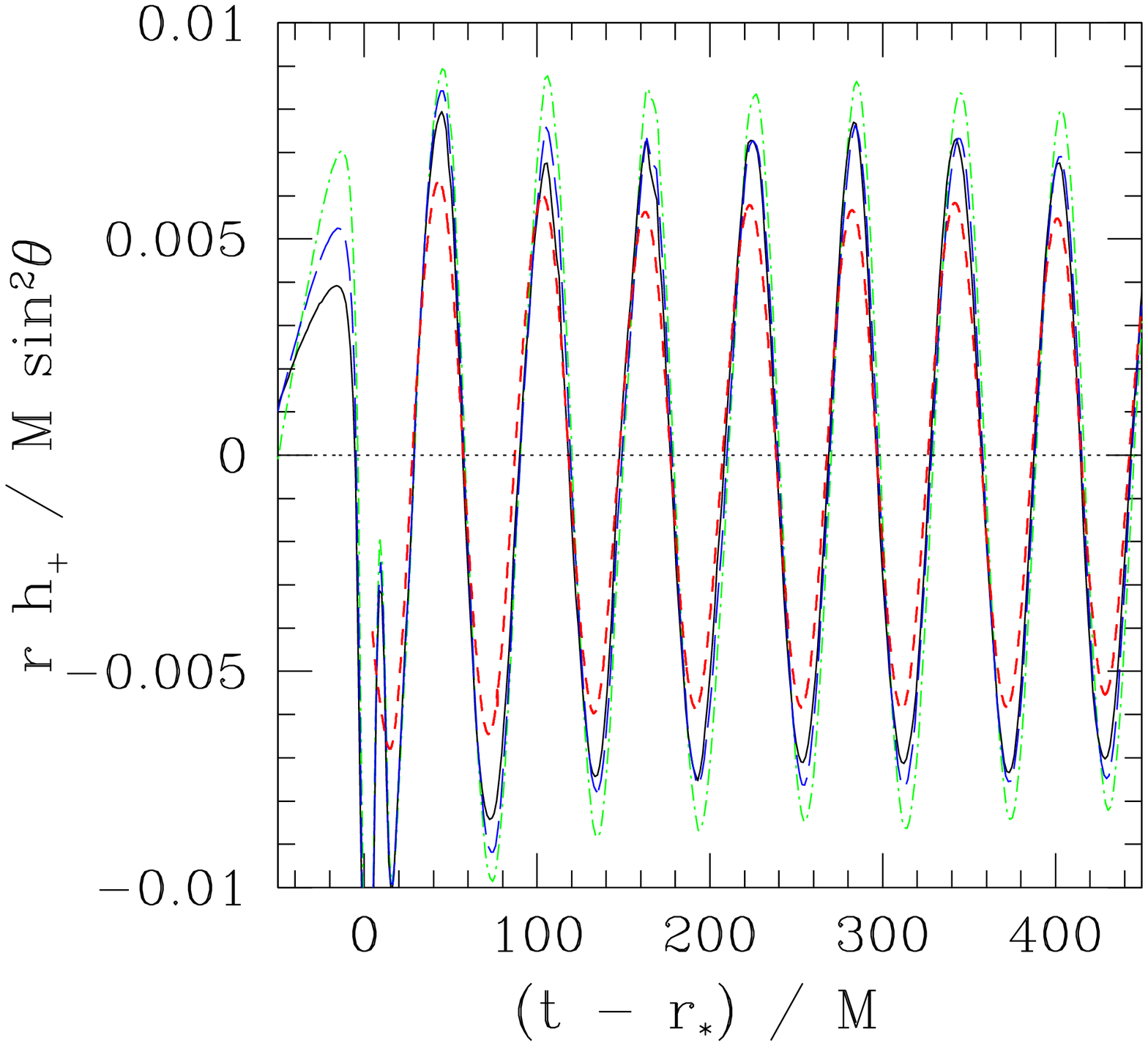}
%\vspace*{-4mm}
\caption{Gravitational waves with $l=2$ from nonrotating neutron stars 
with axisymmetric oscillations (a) for models S2 (left) and (b) S3 (right). 
For both models, we show the results with $N=480$ (dotted-dashed curve),
600 (long-dashed curve), and 720 (solid curve). The dashed curves denote 
the corresponding gravitational waveforms by the quadrupole formula. 
Here, the gravitational waveforms are extracted at $L_{e} \sim 0.9 L$. 
Note that the behavior of the raw data is very similar to that in Fig. 1,
i.e., with the increase of $L_e$ (with the increase of $N$ for
a fixed grid spacing), the amplitude of the drift is larger. 
\label{FIG3}
}
\end{center}
\end{figure}

Figure 3 indicates that one oscillation mode is dominantly excited. 
As a result, the waveforms are well-approximated by a simple 
sine curve. Indeed, the Fourier spectra of
$r h_+^{\rm quad}/\sin^2\theta$ possess 
one sharp peak, and the oscillation periods are $\approx 0.84$, 
0.84, and 0.85 in units of $2\pi\sqrt{R^3/M}$ for models S1, S2, and S3, 
respectively \footnote{In \cite{Font2002}, a three-dimensional simulation
for an oscillating and nonrotating star which is almost the same as model S1
was performed. They computed approximate gravitational waveforms
in a near zone ($\lambda/L \approx 0.12$)  and derived the period
for the fundamental quadrupole mode as $0.83 \time 2\pi \sqrt{R^3/M}$. 
This value agrees with ours in $\sim 1\%$ error. }. 
Thus, irrespective of compactness of the neutron stars, 
the oscillation period is $\approx 0.85 \times 2\pi\sqrt{R^3/M}$. 
This implies that the oscillation is associated with 
the fundamental quadrupole mode, since the coefficient ($\approx 0.85$)
depends very weakly on the compactness of the neutron stars
for a given equation of state \cite{Kojima}. 

From Eq. (\ref{dedt}), the luminosity of gravitational radiation 
as a function of time is computed. Since the luminosity also varies
in a periodic manner, we define an averaged energy flux according to 
\beq
\Big\langle {dE \over dt}\Big\rangle \equiv
{1 \over P_{\rm osc}}\int_{t_0}^{t_0+P_{\rm osc}}{dE \over dt} dt,
\label{dedtav}
\eeq
where $t_0$ is a constant. For models S1--S3, the averaged 
luminosity is $\sim 6 \times 10^{-8}~({\rm in~units~of~} G^{-1}c^5)$.  
Therefore, the energy dissipated in one oscillation period is much smaller
than the total mass energy of the system and the radiation reaction
timescale is much longer than the oscillation period. 

Figures 1 and 3 indicate that 
(i) the wave length is independent of $L$ and $L_e$, 
(ii) the amplitude of gravitational waves is overestimated 
for the smaller values of $L$ and $L_e$, and 
(iii) the amplitude for model S3 approximately converges to an
asymptotic value for $N \agt 600$ within $\sim 10\%$ error. 
These facts suggest that for $L_e \agt \lambda$ 
(see Table II), convergent gravitational waveforms within $10\%$ 
error can be computed. On the other hand, if the outer boundaries are
located in a near zone with $L < \lambda$, the amplitude of
gravitational waves is overestimated: For $L \approx 2\lambda/3$,
it is overestimated by a factor of $\sim 20\%$. 
This result is consistent with that in our previous study 
for gravitational waves from binary neutron stars
in quasiequilibrium circular orbits \cite{SU}. 

The quadrupole formula yields well-approximated gravitational waveforms 
besides a systematic underestimation of the wave amplitude. 
For models S2 and S3, the asymptotic amplitudes of $A_2/M$ are 
about 0.010 and 0.007, respectively. 
On the other hand, according to the quadrupole formula, they are 
about 0.008 and 0.0055. Thus, 
the underestimation factor is $\sim 20$\%. 
If it is due to the first post Newtonian correction, 
the factor should be proportional to the compactness of
neutron stars $M/R(=GM/Rc^2)$ or $v^2(=v^2/c^2)$ 
where $v$ denotes typical magnitude of the oscillation velocity. 
To determine which the dominant component is, 
we performed a simulation reducing the magnitude of 
the velocity perturbation initially given (i.e., reducing 
the magnitude of $V$) and found that the wave amplitude 
is underestimated by $\sim 20$\% irrespective of the 
magnitude of $V$. Therefore, we conclude that 
the underestimation factor is proportional to 
the magnitude of $M/R$ in the present case. 

Although the wave amplitude is underestimated, 
the wave phase is computed accurately by the quadrupole formula. 
The most important element 
in detection of gravitational waves using matched filtering techniques 
is to a priori know the phase of gravitational waves. 
From this point of view, the quadrupole formula is a useful tool
for computation of gravitational waveforms.

\subsubsection{Rotating stars}

In Fig. 4, gravitational waveforms with $l=2$ from oscillating and
rapidly rotating neutron stars for models R2 and R3 are displayed. 
For these simulations, velocity perturbations are added initially 
without changing other quantities. The numerical results are shown 
for $N=480$ (long-dashed curve) and 720 (solid curve).
For $N=480$ and 720, the gauge-invariant variables are 
extracted at $\approx 2\lambda/3$ and $\lambda$, respectively.
The waveforms computed by the quadrupole formula (dashed curves) are
displayed together.

\begin{figure}[htb]
\vspace*{-4mm}
\begin{center}
\epsfxsize=2.9in
\leavevmode
R2\epsffile{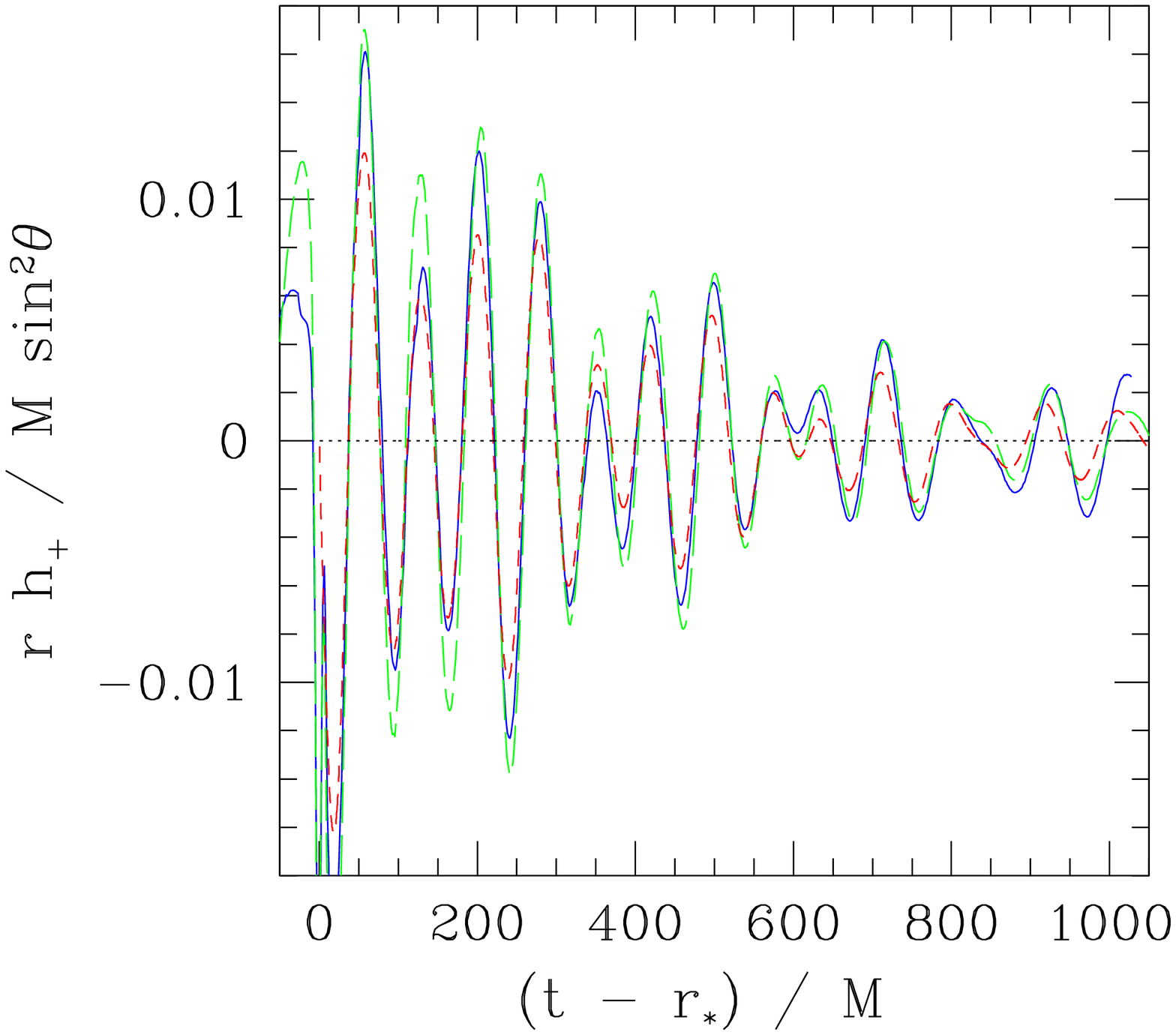}
\epsfxsize=2.9in
\leavevmode
~~~R3\epsffile{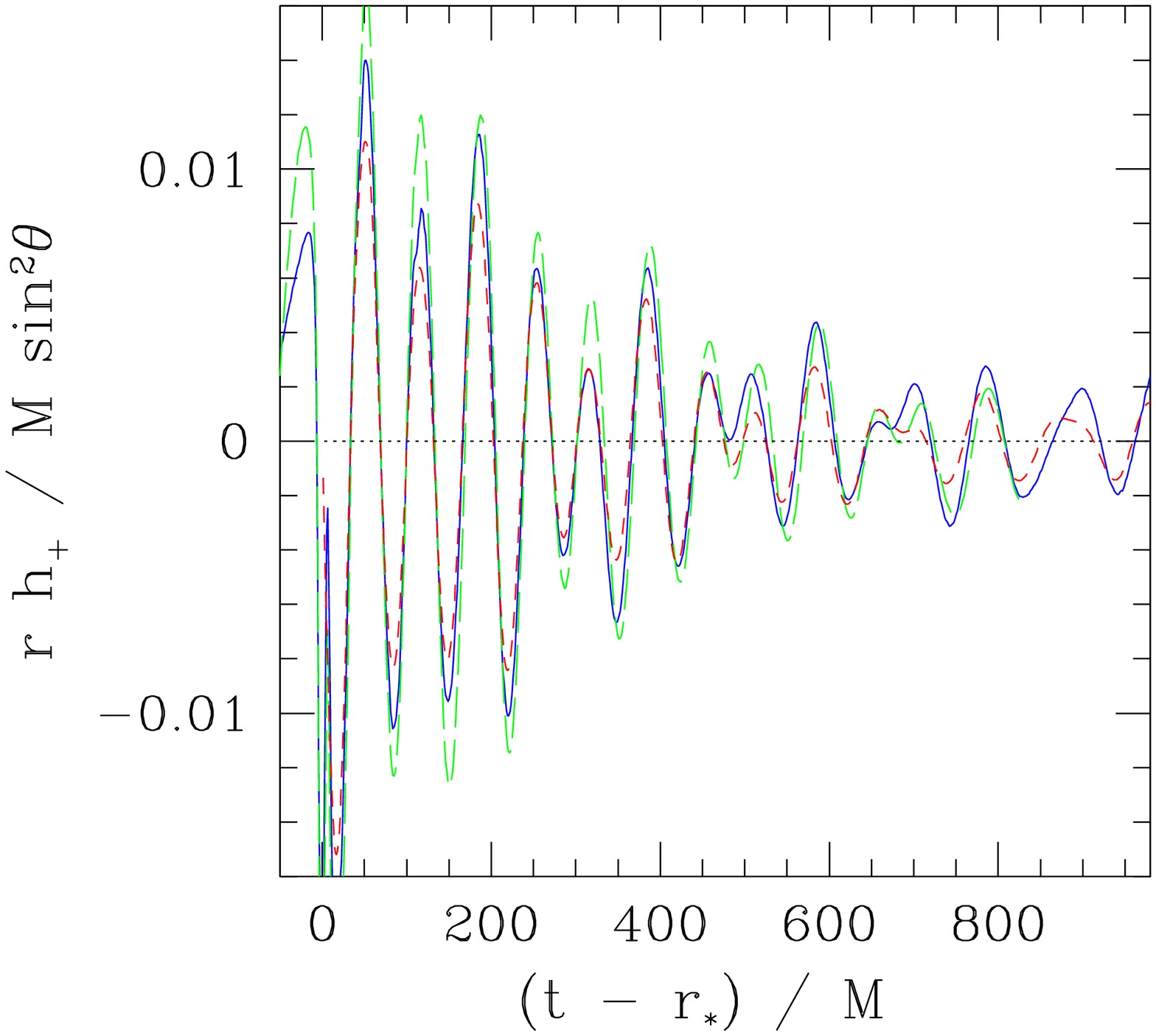}
%\vspace*{-4mm}
\caption{Gravitational waves with $l=2$ from 
rotating neutron stars R2 (left) and R3 (right) 
with $N=480$ (long-dashed curve) and 720 (solid curve).
The dashed curves denote the results by the quadrupole formula. 
\label{FIG4}
}
\end{center}
\end{figure}

\begin{figure}[htb]
\vspace*{-4mm}
\begin{center}
\epsfxsize=2.9in
\leavevmode
R2\epsffile{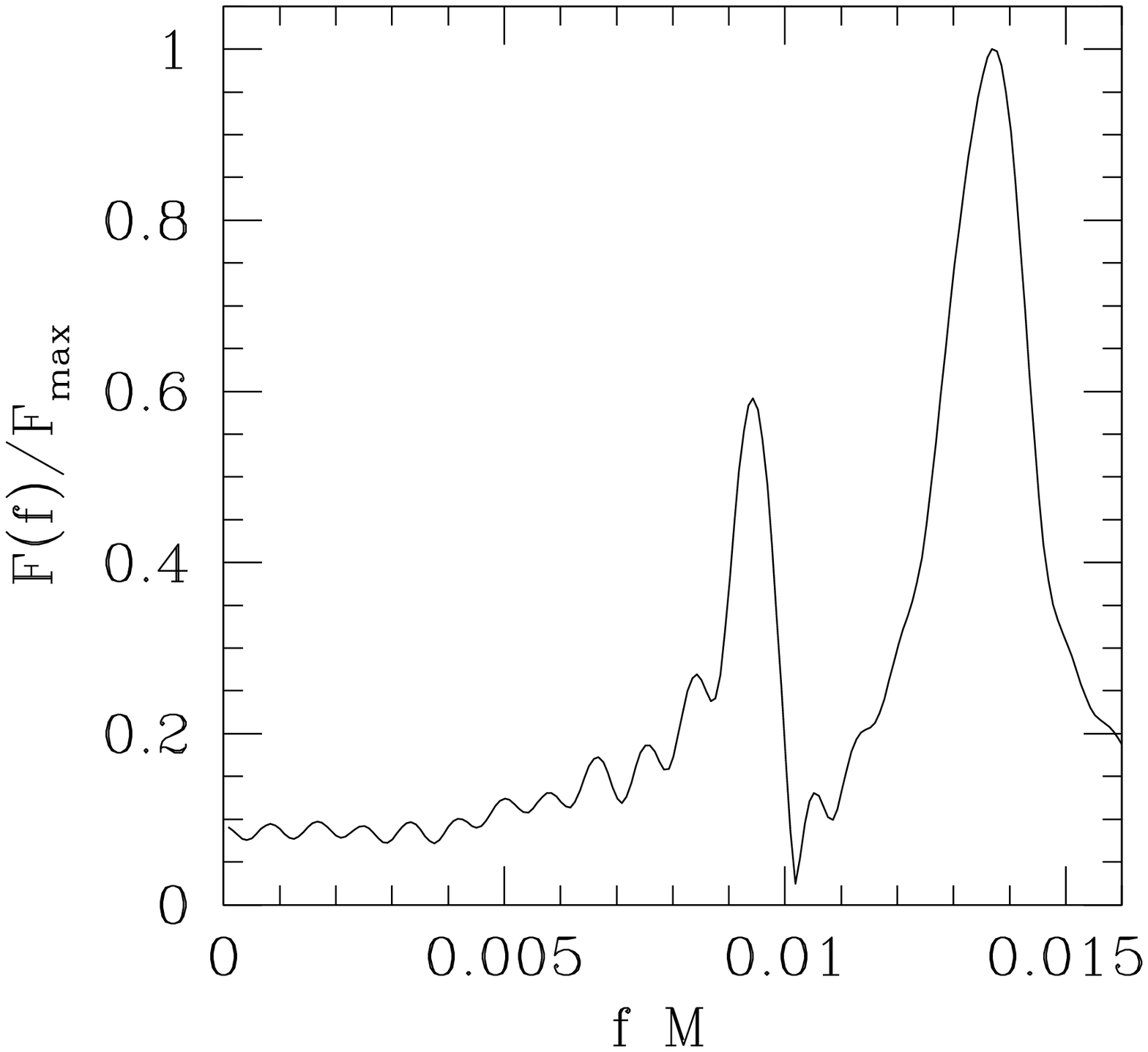}
\epsfxsize=2.9in
\leavevmode
~~~R3\epsffile{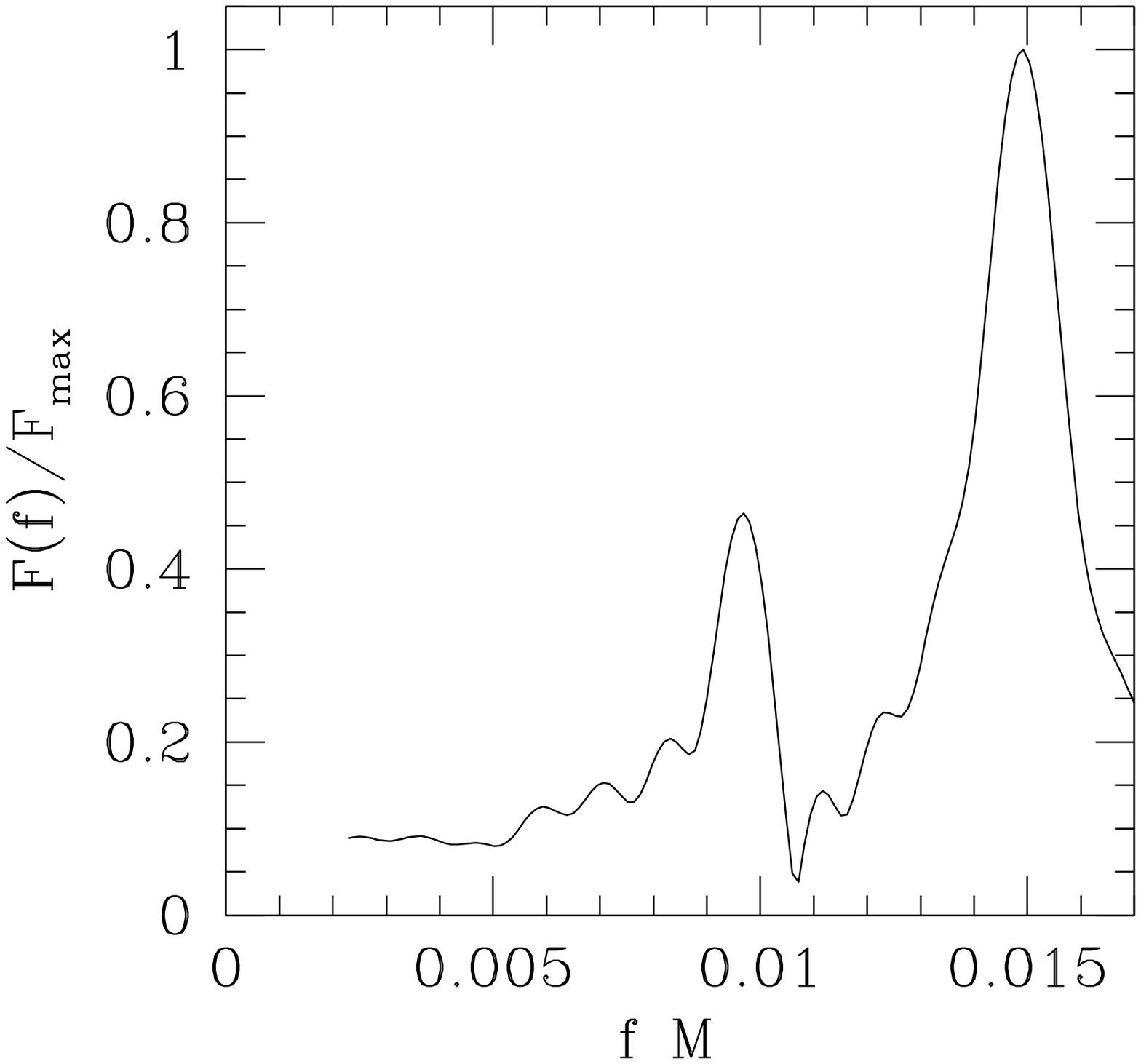}
%\vspace*{-4mm}
\caption{The Fourier spectrum of gravitational waveforms 
for models R2 (left) and R3 (right). 
The spectrum is normalized by the maximum value, and 
the frequency $f$ is shown in units of $M^{-1}$.
The peaks of the smaller and larger frequencies 
indicate the quasiradial and quadrupole modes, respectively. 
\label{FIG5}
}
\end{center}
\end{figure}

In contrast to those from oscillating and nonrotating neutron stars, 
the gravitational waveforms are not of simple sine curve.
There are two main reasons. One is the following: 
In the nonrotating case, the restoring forces 
against a compression for $\varpi$ and $z$ directions are
of identical magnitude and, hence,
the oscillation periods of two directions agree. 
For the rotating stars, on the other hand, the two restoring forces are
not of identical magnitude and, therefore, the oscillation periods
of two directions do not agree. Due to this fact, 
the waveforms are composed of more than two oscillation modes.

The other reason is that gravitational waves are emitted due to
a quasiradial motion in the case of rotating neutron stars in
contrast to nonrotating neutron stars. In particular,
we here choose rapidly rotating neutron stars and, therefore,
the wave amplitude can be
as larger as than for the quadrupole oscillations. 

To clarify what oscillation modes are relevant, 
the Fourier spectrum $F(f)$ for models R2 and R3 are 
displayed in Fig. 5. This shows that
there are two characteristic peaks in the spectrum. 
The oscillation periods defined by $1/f_{\rm peak}$
where $f_{\rm peak}$ is the frequency of the peaks in the Fourier
spectrum are listed in Tables I and III. 

For models R1--R3, the oscillation period of the larger frequency is 
\beq
P_{\rm osc} \approx 0.5\times 2\pi\sqrt{{R^3 \over M}}, \label{eqpr}
\eeq
where $R$ is the equatorial circumferential radius. 
The coefficient ($\approx 0.5$) depends very weakly on the compactness 
of the neutron star. This indicates that the oscillation mode is the 
fundamental {\it quadrupole} mode. 

The frequency of the other peak is smaller than that of the
fundamental quadrupole mode. 
This peak is likely to be associated with the fundamental 
{\it quasiradial} oscillation mode ($p_1$ mode). 
To confirm this prediction, 
we performed the Fourier analysis to the time sequence of the
central density and found that the characteristic oscillation period 
indeed coincides with that of the second peak (see Table III). 
Furthermore, it agrees with the characteristic frequency 
for quasiradially oscillating neutron stars presented
in \cite{S2002}. All these facts confirm our prediction. 

%%It is worthy to note that the quasiradial oscillation
%%of rapidly rotating neutron stars yields
%%gravitational waves of fairly large amplitude. 

The Fourier spectra indicate that 
gravitational waves from axisymmetric global 
oscillations of rigidly rotating stars 
are composed of two dominant modes. To confirm 
this conjecture, we performed simulations initiated from another
initial condition in which 
the pressure is uniformly depleted but the velocity field is unperturbed. 
To uniformly deplete the pressure, $K$ is decreased by 20\% at $t=0$.  
In Fig. 6, we display (a) the gravitational waveform 
and (b) the Fourier spectrum for model R1 for this simulation.
Figure 6 (b) shows that two modes, 
the fundamental quadrupole and quasiradial ones, are dominant again. 
This result justifies our conjecture. In contrast to the cases
in which the nonspherical velocity perturbation is added,
the mode with lower frequency, i.e., the quasiradial mode, is dominant.
This is because the matter motion is almost quasiradial. 

In the collapse of massive rotating stellar core, a protoneutron star 
will be formed. If the progenitor star is not rapidly rotating
and its degree of differential rotation is not high,
the protoneutron star relaxes to a nearly quasistationary state soon 
after the collapse (e.g., \cite{HD}). At the formation of 
a rotating protoneutron star in a nearly quasistationary state,
nonspherical oscillations are excited by the quasiradial infall. 
Because of the nonspherical nature, 
gravitational waves are emitted \cite{HD}.
As illustrated above, in such oscillating neutron stars 
in a nearly quasistationary state, two dominant modes 
(quasiradial and quadrupole modes) may be excited. 
If the collapse is quasiradial, the quasiradial mode will be
the main component. On the other hand,
if the nonspherical quadrupole oscillation of
high amplitude is excited, the quadrupole mode will be dominant. 

\begin{figure}[htb]
\vspace*{-4mm}
\begin{center}
\epsfxsize=2.9in
\leavevmode
(a)\epsffile{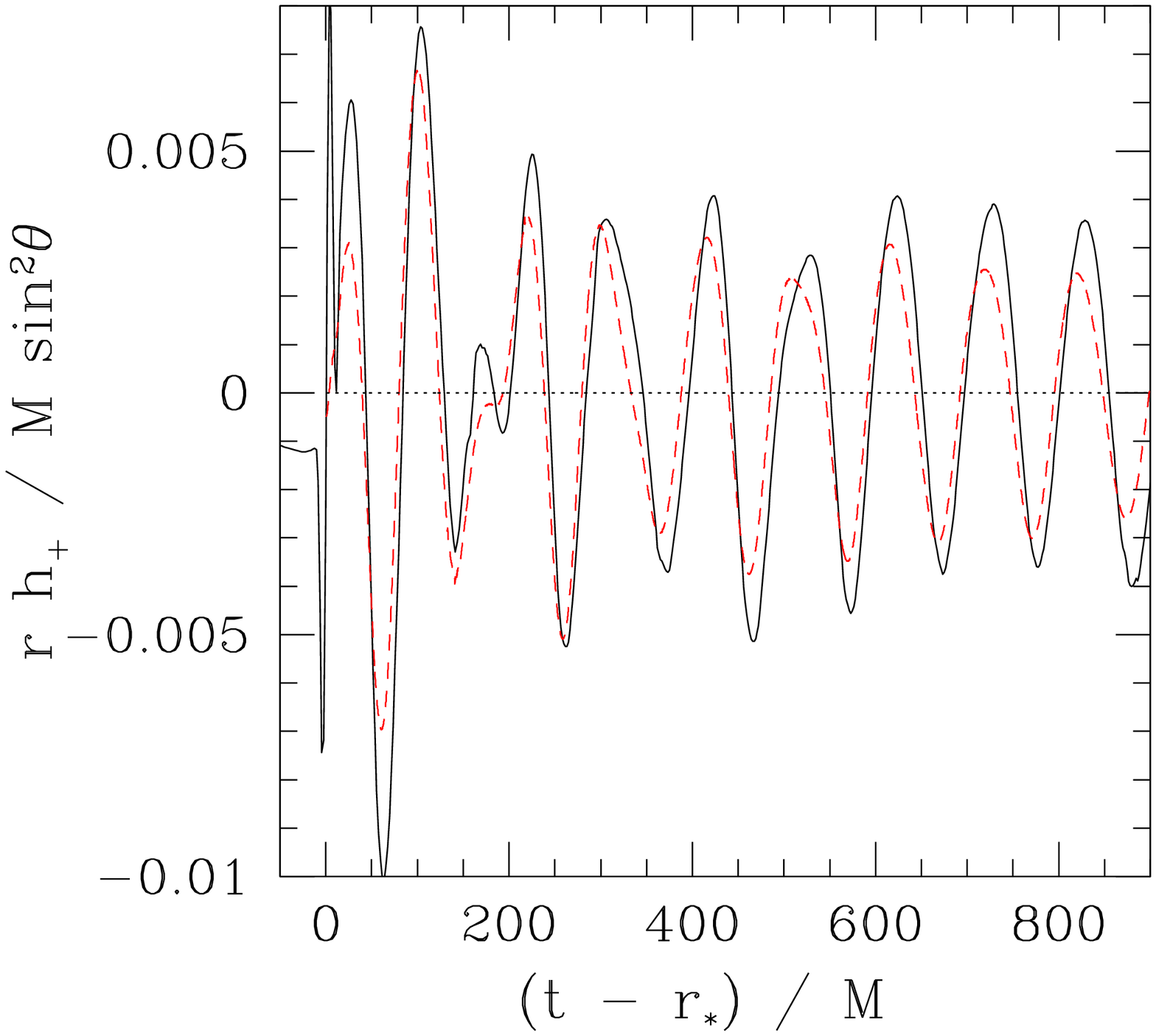}
\epsfxsize=2.9in
\leavevmode
~~~(b)\epsffile{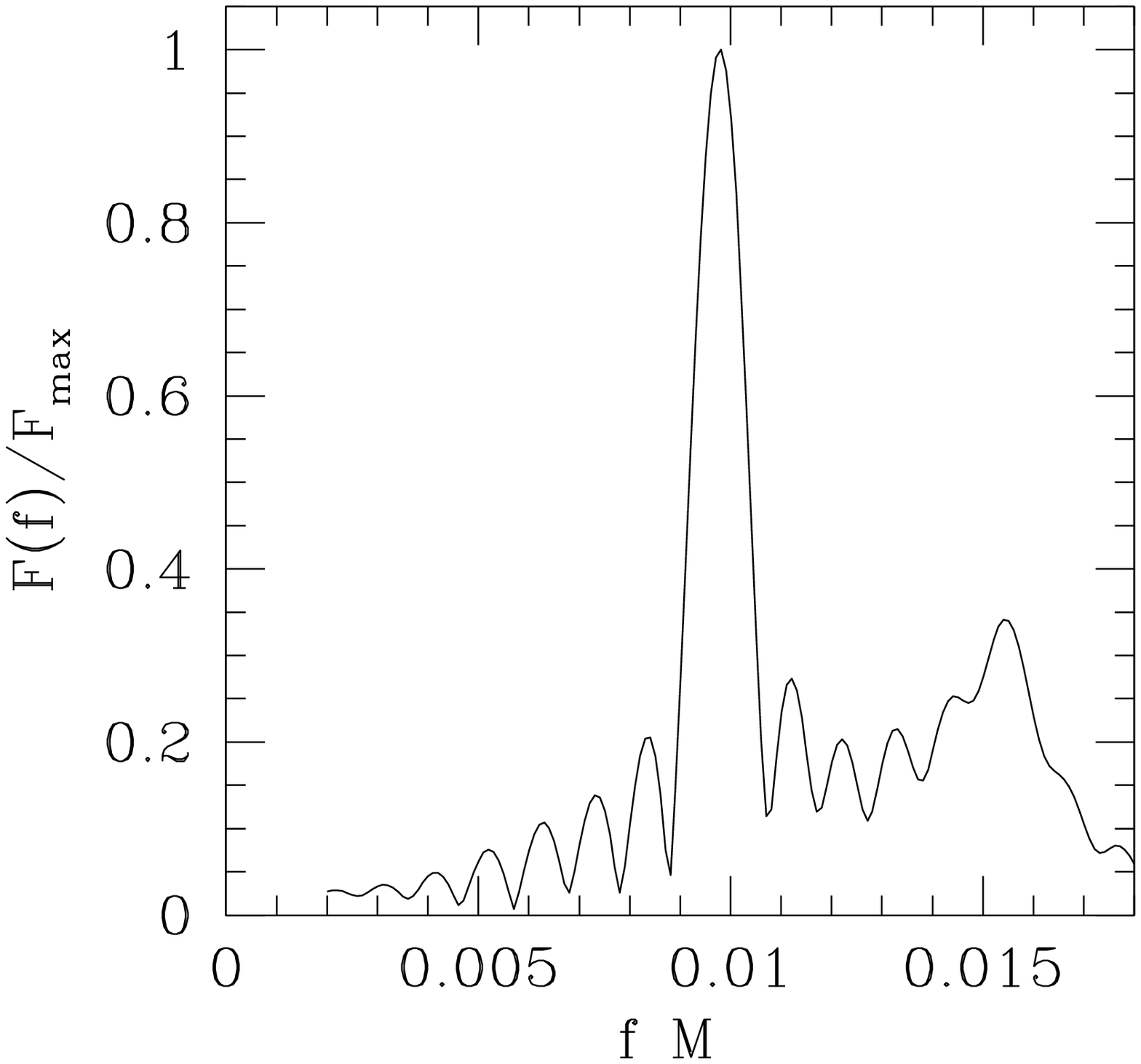}
%\vspace*{-4mm}
\caption{(a) Gravitational waves with $l=2$ from 
rotating neutron stars R1 in a quasiradial oscillation  
with $N=480$ (solid curve). 
The dashed curves denote the results by the quadrupole formula.
(b) The Fourier spectrum of gravitational waves. 
The spectrum is normalized by the maximum value, and 
the frequency $f$ is shown in units of $M^{-1}$. 
The peaks of the smaller and larger frequencies 
indicate the quasiradial and quadrupole modes, respectively.
In this case, the quasiradial mode is dominant (compare with Fig. 5). 
\label{FIG7}
}
\end{center}
\end{figure}

As Fig. 4 shows, the wave amplitude decreases with time
in the early phase ($t-r_* \alt 500M$), and then relaxes to
a small value. 
On the other hand, the amplitude does not decrease in Fig. 6. 
This indicates that the amplitude of the quadrupole oscillation
decreases with time to a small value,
while the quasiradial oscillation does not. 
As mentioned above, 
the oscillation frequencies for $z$ and $\varpi$ directions
are not identical for the quadrupole oscillation of rotating stars.
As a result, shocks may be formed at a collision of 
compression waves in the two different oscillation directions. 
Repeating this process, the quadrupole oscillation may be gradually 
damped. On the other hand, there is no process that damps 
quasiradial mode quickly. Thus, it is reasonable to expect that 
the quasiradial mode eventually dominates 
in the oscillating and rotating stars.

Using Eq. (\ref{dedt}), the luminosity is computed. 
It is found that until $t \sim 1000M$, the total radiated energy 
is computed as $\sim 5 \times 10^{-5} M$, 
which is much smaller than the total mass energy of the system. 
Therefore, the damping timescale of the wave amplitude 
due to gravitational waves is much longer than the oscillation and
rotation periods. 

As in the nonrotating case, approximate gravitational waveforms
were computed using the quadrupole formula.
Figures 4 and 6 indicate that the waveforms agree with those
computed by the gauge-invariant method, 
besides a systematic underestimation of the amplitude. 
The underestimation factor is of order $M/R$ as in the nonrotating cases. 
In the case of rotating stars, a 
modulation of the wave amplitude is outstanding. Using 
the quadrupole formula, however, such modulation can be well 
computed. As mentioned in IV C 1, the most important element 
in detection of gravitational waves using matched filtering techniques 
is to a priori know the phase and modulation of gravitational waves. 
The results here indicates that they
are computed well in the quadrupole formula. 
Therefore, for computation of 
gravitational waves in rotating stellar core collapse 
to a protoneutron star, the quadrupole formula may be a useful tool.

\begin{table}[t]
\begin{center}
\begin{tabular}{|c|c|c|c|} \hline
& $P_{\rm osc}/(2\pi R^{3/2}M^{-1/2})$(for central density) &
$P_{\rm osc}/(2\pi R^{3/2}M^{-1/2})$(for gravitational waves) &
Ref. \cite{S2002} \\ \hline
R1 & 0.66 & 0.66 & 0.66 \\ \hline
R2 & 0.69 & 0.70 & ---  \\ \hline
R3 & 0.76 & 0.76 & 0.76 \\ \hline
\end{tabular}
\caption{
$P_{\rm osc}/(2\pi R^{3/2}M^{-1/2})$ for
the fundamental quasi-radial mode calculated
from the evolution of the central density and
gravitational waveforms.
The last column shows the results obtained from 
simulations of the quasiradial oscillation. 
}
\end{center}
\vspace{-5mm}
\end{table}

\section{Summary and discussion}

We have studied gravitational waves from axisymmetrically
oscillating neutron stars adopting the gauge-invariant wave
extraction method as well as the quadrupole formula. It is found that 
several types of the nonwave components such as the 
stationary parts of metric and numerical errors 
are contained in the gauge-invariant variables.
The numerical errors are generated 
due to an approximate treatment for the outer boundary conditions. 
We illustrate a method to subtract the
dominant components of the numerical errors and 
demonstrate it possible to extract gravitational waves 
even from such noisy data sets with a residual of magnitude $\sim 10^{-3}$.

The gravitational waveforms computed in the quadrupole formula
agree well with those obtained from the gauge-invariant technique
besides a systematic underestimation of the amplitude by $\sim 20$\%. 
An important point is that the evolution of the wave phase and the 
modulation of the amplitude are computed with a good accuracy. 
This indicates that for a study of gravitational waveforms
from rotating stellar core collapse to a protoneutron star, 
the quadrupole formula will be a useful tool in fully relativistic 
simulations. It should be also addressed that 
the result in this paper supports the treatment in \cite{HD} 
in which gravitational waveforms are computed using 
a quadrupole formula in approximate general relativistic simulations. 

The gauge-invariant variables are extracted 
for various values of extraction radii.  
It is found that to extract gravitational waves 
within $\sim 10\%$ error, the extraction radius has to 
be larger than $\sim 90\%$ of the gravitational wave length. 
If the outer boundaries are located in the near zone with $L < \lambda$,
the amplitude of gravitational waves is overestimated:
For $L \approx 2\lambda/3$, it is overestimated by $\sim 20\%$. 
For $L < 2\lambda/3$, the factor of the overestimation is even larger.

In the present work, the amplitude of gravitational waves in a
local wave zone is much larger than that of systematic numerical errors. 
This fact enables to subtract them 
from the gauge-invariant variables accurately.
If the magnitude of the errors is much larger than that
of the amplitude of gravitational waves, however, 
it would not be possible to carry out an accurate subtraction. 
For example, in rotating stellar core collapse, the
amplitude of gravitational waves in the local wave zone at $r \sim \lambda$
is at most $\sim 10^{-5}$ according to gravitational waveforms 
calculated by a quadrupole formula \cite{HD}. 
To extract gravitational waves of such small amplitude, 
it is necessary to reduce the magnitude of the numerical errors. 
To achieve that, we need to impose more accurate outer boundary conditions. 
Developing such conditions is crucial in computing gravitational waves
of small amplitude of $O(10^{-5})$ from raw data sets of metric. 

Another possible method for computing accurate gravitational 
waves of small amplitude is to adopt a quadrupole formula taking
into account higher-order post Newtonian terms.
As indicated in this paper, the simple quadrupole formula 
underestimates the amplitude of gravitational waves by $O(M/R)$.
In rotating stellar core collapse, the error in the amplitude will be
$\sim 10\%$. To compute the amplitude within $\sim 1\%$ error,
we should take into account higher general relativistic corrections. 
In quadrupole formulas with the higher post Newtonian 
corrections (as derived in \cite{BDS}), 
it may be possible to obtain gravitational waveforms within
1\% error. Such formulas will be useful to extract gravitational waves 
of small amplitude from rotating stellar core collapses and from
oscillating neutron stars. 

In addition to the study for gravitational wave extraction, 
oscillation modes of rotating neutron stars are analyzed.
It is found that two modes 
(the fundamental quadrupole and quasiradial modes) 
are dominantly excited due to the global oscillation. 
The frequency of the quadrupole mode is proportional to $\sqrt{GM/R^3}$, 
and is higher than that of the quasiradial one for the typical values 
of mass and radius of neutron stars. 
It is shown that the amplitude of the quadrupole mode decreases 
with time due to an incoherent nature of the oscillation, but 
that of the quasiradial mode is not damped quickly, 
hence being the dominant mode after several dynamical timescales. 
We expect that in rotating stellar core collapse 
to a protoneutron star in a nearly quasistationary state, 
these two modes may be the main components in the 
burst phase of gravitational waves. 
The quadrupole mode will be damped within a few dynamical 
timescales and subsequently the quasiradial mode will be the
dominant component to be longterm quasiperiodic waves. 

\begin{center}
{\bf Acknowledgments}
\end{center}

Numerical computations were carried out 
on the FACOM VPP5000 machine in the data processing center of
National Astronomical Observatory of Japan.
This work is in part supported by Japanese Monbu-Kagakusho Grant
(Nos. 13740143, 14047207, 15037204, and 15740142). 

\appendix

\section{Solutions for the tensor Laplace equation}

Here, we describe solutions for the tensor Laplace
equations in spherical polar coordinates as 
\beq
\Delta_f h_{ab}=0.
\eeq
$h_{ab}$ is expanded by tensor harmonic functions as 
\beqn
h_{ab}=&& \sum_l \left(
\begin{array}{lll}
\displaystyle 
A_{l} Y_{l0} & r B_{l} Y_{l0,\theta}& 0\\
\ast& r^2(K_{l}Y_{l0}+G_{l}W_{l0}) & 0 \\
\ast& \ast&r^2\sin^2\theta(K_{l}Y_{l0}-G_{l}W_{l0}) \\
\end{array}
\right) \nonumber \\
&&+\left(\begin{array}{ccc}
0 &  0 & r C_{l} \pa_{\theta}Y_{l0}\sin\theta  \\
\ast & 0 & -r^2 D_{l}W_{l0}\sin\theta  \\
\ast & \ast & 0 \\
\end{array}
\right),
\eeqn
where $A_l$, $B_l$, $C_l$, $D_l$, $G_l$, and $K_l$ are
functions of $r$. With the above expansion, 
the components of the Laplace equation are written as
\beqn
&& \Delta_f h_{rr}= \sum_{l} 
\biggl[ A_{l}'' + {2 \over  r} A_{l}'-{\lambda_l +4 \over r^2} A_{l}
+{4 \over r^2} K_l+{4\lambda_l \over r^2} B_{l}\biggr]Y_{l0}=0, \nonumber \\
%%%%%%%%%%%%%%%%%%%
&& \Delta_f h_{r\theta}=r \sum_{l} 
\biggl[ B_{l}'' + {2 \over r} B_{l}'- {\lambda_l +4 \over r^2} B_{l}
+{2 \over r^2} A_{l}-{2 \over r^2} K_{l}
+{2\lambda_l -4 \over r^2} F_{l}
\biggr] \pa_{\theta} Y_{l0}=0, \nonumber \\
%%%%%%%%%%%%%%%%%%%%%%%%%%%%%%
&& \Delta_f h_{r\varphi}=r \sum_{l} 
\biggl[ C_{l}'' +{2 \over r} C_{l}' - {\lambda_l +4 \over r^2} C_{l}
+{-2\lambda_l + 4 \over r^2} D_{l}
\biggr]{\pa_{\varphi} Y_{l0} \over \sin\theta}=0, \nonumber \\
%%%%%%%%%%%%%%%%%%%%
&& {\Delta_f h_{\theta\varphi} \over  r^2} = \sum_{l} 
\biggl[ D_{l}'' + {2 \over r} D_{l}' - {\lambda_l -2 \over r^2} D_{lm}
- {2 \over r^2} C_{l} \biggr] \sin\theta W_{l0}=0, \nonumber \\
%%%%%%%%%%%%%%%%%%%%
&& {\Delta_f h_{\theta\theta} \over r^2} 
= \sum_l
\biggl[\biggl( K_l'' + {2 \over r} K_l' - {\lambda_l+2 \over r^2}K_l
+{2 \over r^2} A_l -{2\lambda_l \over r^2} B_l\biggr)Y_{l0} 
+\biggl( F_l'' + {2 \over r}F_l' -{\lambda_l-2 \over r^2} F_l
+{2 \over r^2}B_l \biggr) W_{l0}\biggr] \nonumber \\
&&~~~~~~~~\equiv \sum_{l}
\biggl(H^k_{l}Y_{l0} + H^f_{l}W_{l0} \biggr)=0,\nonumber\\
%%%%%%%%%%%%%%%%%%%%
&& {\Delta_f h_{\varphi\varphi} \over r^2\sin^2\theta} 
=\sum_{l}\biggl(-H^k_{l}Y_{l0} - H^f_{l}W_{l0}\biggr)=0,
\eeqn
where $\lambda_l=l(l+1)$ and $'$ denotes $d/dr$.

Setting $A_l=\bar A_l r^n$, $B_l=\bar B_l r^n$, $K_l=\bar K_l r^n$, 
and $F_l=\bar F_l r^n$ where $\bar A_l \sim \bar F_l$ are constants,
simultaneous equations for the even-parity modes are derived as
\beqn
\left(
\begin{array}{llll}
x-4 & 4\lambda_l & 4 & 0 \\
2 & x-4 & -2 & 2\lambda_l-4 \\
2 & -2\lambda_l & x-2 & 0\\
0 & 2 & 0 & x+2 \\
\end{array}
\right)
\left(
\begin{array}{l}
\bar A_l \\
\bar B_l \\
\bar K_l \\
\bar F_l \\
\end{array}
\right)=0,
\eeqn
where $x=n(n+1)-\lambda_l$. For the existence of the solutions, 
the determinant of the matrix should be zero. Then, 
an algebraic equation for $n$ is derived, and the solutions are
$n=l\pm 2$, $l$, $-l \pm 1$ and $-l-3$. 
The relations among $\bar A_l \sim \bar F_l$ are easily obtained
for each value of $n$. It is also easy to check that 
the solutions with $n=l-2$ and $-l-3$ are written
in the form $V_{i|j}+V_{j|i}$ using a vector $V_i$ of even-parity. 

From the same procedure, the solutions for the odd-parity mode 
are written as $C_l=\bar C_l r^{n}$ and $D_l=\bar D_l r^{n}$ 
where $n=l \pm 1$, $-l$ and $-l-2$, and $\bar C_l$ and $\bar D_l$
are constants. In this case, 
the solutions with $n=l-1$ and $-l-2$ are written as $V_{i|j}+V_{j|i}$ 
using a vector $V_i$ of odd-parity.


\begin{thebibliography}{99}

\bibitem{HD} H. Dimmelmeier, J. A. Font and E. M\"uller,
Astron. Astrophys. {\bf 388}, 917 (2002); {\bf 393}, 523 (2002). 

\bibitem{Siebel} F. Siebel, J. A. Font, E. M\"uller and P. Papadopoulos,
Phys. Rev. D {\bf 67}, 124018 (2003).

\bibitem{Newton} L. S. Finn and C. R. Evans, Astrophys. J. {\bf 351}, 
588 (1990).

\bibitem{Newton1} R. M\"onchmeyer, G. Sch\"afer, E. M\"uller and R. 
Kates, Astron. and Astrophys. {\bf 246}, 417 (1991): 
E. M\"uller and H.-T. Janka, Astron. Astrophys. {\bf 103}, 358 (1997).

\bibitem{Newton2}
S. Bonazzola and J.-A. Marck, Astron. Astrophys. {\bf 267}, 623 (1993). 

\bibitem{Newton3}
S. Yamada and K. Sato, Astrophys. J. {\bf 434}, 268 (1994):
{\bf 450}, 245 (1995): K. Kotake, S. Yamada, and K. Sato, Phys. Rev. D
{\bf 68}, 044023 (2003).

\bibitem{Newton4} T. Zwerger and E. M\"uller, Astron. Astrophys.
{\bf 320}, 209 (1997): 
M. Rampp, E. M\"ulelr and M. Ruffert, Astron. Astrophys.
{\bf 332}, 969 (1998). 

\bibitem{Newton5}
C. Fryer and A. Heger, Astrophys. J. {\bf 541}, 1033 (2000):
C. Fryer, D. E. Holz and A. Heger, Astrophys. J. {\bf 565}, 430
(2002). 

\bibitem{Newton6} C. D. Ott, A. Burrows, E. Livne, and R. Walder,
Astrophys. J. {\bf 600}, 834 (2004).

\bibitem{AE} 
A. M. Abrahams and C. R. Evans, Phys. Rev. D {\bf 37},
318 (1988); {\bf 42}, 2585 (1990). 

\bibitem{Siebel0} F. Siebel, J. A. Font, E. M\"uller and P. Papadopoulos,
Phys. Rev. D {\bf 65}, 064038 (2002). 

\bibitem{S2002} M. Shibata, Phys. Rev. D {\bf 67}, 024033 (2003). 

\bibitem{alcu} M. Alcubierre, S. Brandt, B. Br\"ugmann, 
D. Holz, E. Seidel, R. Takahashi and J. Thornburg,
Int. J. Mod. Phys. D {\bf 10}, 273 (2001). 

\bibitem{Font}
J. A. Font, Living Review Relativity {\bf 3}, 2, 2000
http://www.livingreviews.org/Articles/Volume2/2000-2font: F. Banyuls, 
J. A. Font, J.-Ma. Ib\'a\~nez, J. M. Marti, and J. A. Miralles, 
Astrophys. J. {\bf 476}, 221 (1997). 

\bibitem{Font2002}
J. A. Font, T. Goodale, S. Iyer, M. Miller, L. Rezzolla, E. Seidel,
N. Stergioulas, W. M. Suen and M. Tobias, Phys. Rev. D {\bf 65}, 084024 
(2002).

\bibitem{shibata} M. Shibata, Prog. Theor. Phys. {\bf 101}, 1199 (1999); 
Prog. Theor. Phys. {\bf 104}, 325 (2000).

\bibitem{gr3d}
M. Shibata, Phys. Rev. D {\bf 60}, 104052 (1999). 

\bibitem{SBS} M. Shibata, T. W. Baumgarte and 
S. L. Shapiro, Phys. Rev. D {\bf 61}, 044012 (2000); 
Astrophys. J. {\bf 542}, 453 (2000). 

\bibitem{bina} M. Shibata and K. Ury\=u, Phys. Rev. D {\bf 61}, 064001
(2000): Prog. Theor. Phys. {\bf 107}, 265 (2002). 

\bibitem{SN} In \cite{shibata,SBS,bina} and this paper,
we adopt a formulation slight modified from the original 
version presented in the following reference:  
M. Shibata and T. Nakamura, Phys. Rev. D {\bf 52}, 5428 (1995). 
See also T. Nakamura, K. Oohara, and Y. Kojima, 
Prog. Theor. Phys. Suppl. {\bf 90}, 1 (1987). 

\bibitem{S03} M. Shibata, Astrophys. J. {\bf 595}, 992 (2003). 

%\bibitem{SS} M. Shibata and S. L. Shapiro, Astrophys. J. Lett. {\bf 572},
%L39 (2002). 

\bibitem{moncrief} V. Moncrief, Ann. of Phys. {\bf 88}, 323  
(1974)

\bibitem{Abrahams} A. Abrahams, D. Bernstein, D. Hobill, E. Seidel and L. Smarr,
  Phys. Rev. D {\bf 45}, 3544 (1992).

\bibitem{T82} S. A. Teukolsky, Phys. Rev. D {\bf 26}, 745 (1982). 

\bibitem{CST} G. Cook, S. L. Shapiro, and S. A. Teukolsky, 
Astrophys. J. {\bf 422}, 227 (1994). 

%\bibitem{foot1} 

%\bibitem{foot2} 

\bibitem{Kojima} Y. Kojima, Prog. Theor. Phys. {\bf 77}, 297 (1987). 

\bibitem{N84} T. Nakamura, Prog. Theor. Phys. {\bf 72}, 746 (1984). 

\bibitem{SU} M. Shibata and K. Ury\=u, Phys. Rev. D {\bf 64},
104017 (2001).

\bibitem{BDS} L. Blanchet, T. Damour and G. Sch\"afer,
Mon. Not. R. astr. Soc. {\bf 242}, 289 (1990). 


\end{thebibliography}
\end{document}